\documentclass[review]{elsarticle}
\usepackage{amsthm}
\usepackage{amssymb}
\usepackage{algorithm}
\usepackage{algorithmic}
\usepackage{diagbox}
\usepackage{lineno,hyperref}
\usepackage{amsmath}
\usepackage{graphicx}
\usepackage{array}
\usepackage{booktabs}
\usepackage{multirow}
\usepackage{bm}
\usepackage{tikz}
\usepackage{fancyhdr}
\renewcommand{\algorithmicrequire}{\textbf{Input:}} 
\renewcommand{\algorithmicensure}{\textbf{Output:}} 

\journal{Journal of Information processing \& management}

\begin{document}

\begin{frontmatter}

\title{SPChain: Blockchain-based Medical Data Sharing and Privacy-preserving eHealth System}

\author[a]{Renpeng Zou \fnref{1}}

\author[a]{Xixiang Lv\corref{mycorrespondingauthor}\fnref{1}}
\cortext[mycorrespondingauthor]{Corresponding author}
\ead{xxlv@mail.xidian.edu.cn}

\author[a]{Jingsong Zhao  \fnref{1}}

\fntext[1]{The work of Xixiang Lv, Renpeng Zou and Jingsong Zhao is supported by the National Natural Science Foundation of China (Grant No. 62072356).}

\address[a]{School of Cyber Engineering, Xidian University,Xian 710071, China}

\begin{abstract} 
The sharing of electronic medical records (EMRs) has great positive significance for research on disease and epidemic prevention. Recently, blockchain-based eHealth systems have achieved great success in terms of EMRs sharing and management, but there still remain some challenges. Permissioned blockchain-based solutions provide high throughput and scalability, but may suffer from rollback attacks and lead to privacy leakage. Designs based on the public blockchain, on the other hand, are more open and secure, but sacrifice scalability and have no incentives for medical institutions to join into the systems. Moreover, data retrieval in blockchain-based eHealth systems is inefficient because of the basic blockchain structure. To solve the above problems, we propose a blockchain-based medical data sharing and privacy-preserving eHealth system named SPChain. To achieve quick retrieval, we devise special keyblocks and microblocks for patients to store their EMRs. A reputation system is also constructed to motivate medical institutions to participate in SPChain. By using proxy re-encryption schemes, SPChain achieves medical data sharing for patients in a privacy-preserving manner. To evaluate SPChain, we leverage the distribution of miners in the real world to test the system's performance and ability to resist mentioned attacks. The results show that SPChain can achieve high throughput (220 TPS) with low storage overhead. Compared with the existing schemes, SPChain achieves lower time complexity in terms of data retrieving, and can resist proposed blockchain attacks as well as SPChain attacks.
\end{abstract}


\end{frontmatter}


\section{Introduction}
The era of big data brings new opportunities and challenges to the medical field. Recently, more and more infectious diseases, such as COVID-19 \cite{fauci2020covid}, have drawn our attentions due to their strong infection and destructive ability. Researchers have found that the sharing of medical data can prevent the large-scale spread of diseases to some extent \cite{jin2019review}. Through the analysis and sharing of medical data, we can detect the symptoms of disease transmission in advance and control the large scale spread of diseases in time. On the other hand, some research institutions need large numbers of reliable samples to study the incidence rate of disease. i.e. artificial intelligence techniques are utilized on diagnostics in some intractable diseases such as glaucoma, hyperactivity, and Parkinson's disease \cite{hirschauer2015computer}. Therefore, medical data sharing is of great significance to the development of human health.

As an indispensable tool for medical services \cite{liu2020b4sdc}, electronic medical records (EMRs) not only provide medical data for diagnosis and scientific research but also give one kind of judgment basis for handling medical disputes. EMRs have become the most original record of patients in the whole process of medical treatment, but the current EMR data management systems are not perfect \cite{hardin2021amanuensis}. EMRs are usually stored in a private database, which brings a problem that patients leave data scattered across various medical institutions because life events take the patients away from one medical institution into another. What's worse, since medical data is recorded into EMRs in hospitals or medical institutions after diagnosis, patients lose easy access to these data even if it belongs to them \cite{campanile2021designing}. 

To address these problems, researchers leverage cloud platform to manage EMRs \cite{yang2015hybrid}. To a certain extent, the application of cloud technology has promoted the sharing of EMRs, but there still remain some problems. Considering the fact that users and cloud providers usually belong to different administrative domains, the difficulty of cloud-based data sharing lies in how much trust users can place on cloud service providers. Such a lack of trust stems from the lack of transparency and the loss of data control \cite{ren2012security} by users in cloud environments. e.g. in recent years, many accidents about medical-record leakages \cite{zhao2020blockchain}, \cite{liu2021sedid}, \cite{liu2016privacy} have occurred frequently. Besides, the multi-tenancy characteristic of public cloud services decides that virtual machines (VMs) are shared among various applications, which may expose the data to different types of attacks. Worse still, it is difficult to detect or monitor such attacks in a shared VM environment. Therefore, eHealth systems crave innovations to assure the security and privacy of medical records.

Blockchain, which is widely leveraged in cyptocurrency systems \cite{nakamoto2019bitcoin}, \cite{casino2019systematic}, is a promising technology that can be used to maintain a transparent ledger and share data among participants. With the tamper-resistant and distributed nature, the blockchain technology can provide integrity and restoration guarantees for medical records. Many countries have combined the blockchain technology with eHealth systems and achieved great success. For instance, Estonia \cite{sullivan2017residency} makes use of the blockchain technology to provide patients with safer and more convenient medical services. An ample amount of solutions, i.e. \cite{jing2021blockchain}, \cite{esposito2021blockchain}, \cite{de2020blockchain}, \cite{bernabe2019privacy}, \cite{berdik2021survey} strive to leverage the latest technologies, such as smart contracts and privacy protection modules, to enhance the operability and confidentiality of eHealth systems. Through the literature review, it would be an effective way to use blockchain to serve as a decentralized storage platform and replace the central servers, but there still remain some drawbacks.

Currently, blockchain-based eHealth systems can be divided into two categories, permissioned blockchain-based eHealth systems and public blockchain-based eHealth systems \cite{jin2019review}, \cite{gordon2018blockchain}. Permissioned blockchain-based eHealth systems (e.g. \cite{xia2017bbds}, \cite{xia2017medshare} and \cite{liang2017integrating}) rely on some super nodes to manage the storage and sharing of EMRs. In spite of the high throughput, the permissioned blockchain is far from a perfect solution for secure medical data sharing because it adopts central authorities which are comprised of a group of companies with a shared interest that will oversee the whole system. Therefore, the immutability of data in permissioned blockchain is discounted, which increases the possibility of a central authority rolling back the blockchain records. 

Designs based on public blockchain, on the other hand, provide stronger security and openness, but sacrifice scalability. The public blockchain is cryptocurrency driven, which means a certain amount of cryptocurrencies have to be paid for transaction inclusion and block mining. Such systems are suitable for cleaning institutions such as banks, but have no incentive for medical institutions. Another obstacle to the development of public blockchain-based eHealth systems is the low efficiency of data retrieval. For the data stored in the general database, we can search directly. But in the blockchain, we need to search the block first, and then search the needed transactions contained in the block. These systems utilize smart contracts and the scripting language to exchange medical data between medical institutions with blockchain systems. Due to the data storage and exchange model, the medical data is scattered in blockchain systems. In this case, consulting the medical records of patients in a mass of block data is inefficient. And obtaining a patient's whole medical records in such systems consumes a lot of time. Worse still, public blockchain-based eHealth systems are inefficient in processing transactions. Taking Bitcoin and Ethereum as examples, the throughput of Bitcoin is 7 TPS (transactions per second) and that of Ethereum is 15 TPS \cite{hu2021transaction}, \cite{xu2021latency}, which is far from enough to meet the urgent needs of patients.

In terms of the above problems, we propose a public blockchain-based eHealth system with high throughput to achieve medical data sharing in a privacy-preserving manner. In order to provide efficient retrieval for patients, we design new block and chain structures  with chameleon hash functions. We also devise new incentive mechanism for medical institutions to join into SPChain. Specifically, the contributions of this paper are as follows.
\begin{itemize} 
\item We propose a data sharing and privacy-preserving public blockchain system named SPChain. We use chameleon hash functions to design new block structures which provide whole medical history for each patient. We also devise new chain structure to mitigate forking problem. To motivate medical institutions to participate in SPChain, we construct a reputation system with which the reputable medical institutions can acquire EMRs for medical research.
\item  We use SPChain to construct a concrete eHealth system which can achieve secure data sharing and retrieval in a privacy-preserving manner. We use the proxy re-encryption schemes to protect the EMRs stored in medical institutions' local databases. Only authorized medical institutions can access patients' EMRs. Besides, the system provides special transactions for patients to register in SPChain and label wrong medical records.
\item We discuss how the proposal can satisfy the security requirements and demonstrate the feasibility and effectiveness of SPChain by developing the system in an analog network with the miner distribution in the real world. Compared with existing systems, SPChain  achieves high throughput with minimal storage overhead, and provides significant resilience to Blockchain attacks as well as SPChain attacks.
\end {itemize}

The remaining part of the paper is organized as follows. We begin by introducing some related works in Section 2. Section 3 is concerned with some preliminaries used in this paper. In Section 4, we describe the system model and design goals. The detailed blockchain-based medical data sharing and privacy-preserving eHealth system is given in Section 5. Section 6 illustrates the performance evaluation and security analysis of SPChain. Finally, this paper is concluded in Section 7.
\section{Related Work}
In this section, we review some research trends about medical data sharing via cloud service and blockchain technology.
 
In order to solve the problem of low efficiency and poor scalability of traditional eHealth systems, some solutions (e.g. \cite{popa2011enabling}, \cite{kumbhare2012cryptonite}, \cite{li2020blockchain} and \cite{yang2015hybrid}) are dedicated to use access control technologies to manage the EMRs outsourced to clouds. Such mechanisms are designed to protect the security of remotely
stored data in cloud computing, which demonstrates that providing owners with data access control is more important than letting the cloud take the full control over their data. However, patients still lose control of their data. Moreover, the distrust of the cloud and the compatibility of various cloud devices still restrict the development of cloud-based eHealth systems.

Recently, with the adoption of blockchain technology becoming a widespread trend in distributed computing, many researchers consider using blockchain to provide secure medical data sharing and management. Blockchain-based eHealth systems can be divided into two types \cite{jin2019review}: permissioned blockchain-based approaches and public blockchain-based approaches. 

\textbf{Permissioned blockchain-based eHealth systems (\cite{xia2017medshare}, \cite{xia2017bbds}, \cite{fan2018medblock}, \cite{zhang2018towards} and \cite{huang2020blockchain}).} To preserve patients' privacy in the process of disseminating EMRs, Xia et la. \cite{xia2017medshare} designed a system that addresses the issue of medical data sharing among medical big data custodians in a trust-less environment. The system employed smart contracts and an access control mechanism to effectively trace behaviors on the data, and revoked access to violated rules and permissions on data. With a data custodian system, the system can monitor entities that access data for malicious behaviors. The author also proposed BBDS \cite{xia2017bbds}, a similar blockchain-based framework that provides data provenance, auditing, and control in cloud repositories among healthcare providers. However, their secure sharing of sensitive medical information is limited to invited and verified users.

Fan et al. proposed MedBlock \cite{fan2018medblock}, a hybrid blockchain-based architecture to secure EMRs. In MedBlock nodes are divided into endorsers, orderers and committers. Its consensus protocol is a variant of the PBFT \cite{castro2002practical} consensus protocol. However, the authors did not explicitly explain the access control policies to allow third-party researchers to access medical data.  

Zhang and Lin \cite{zhang2018towards} designed a hybrid blockchain-based secure and privacy-preserving (BSPP) EMRs sharing scheme, where a private blockchain is used to store EMRs for each hospital and a consortium blockchain is used to keep secure indices of the EMRs. In their design, a public encryption-based keyword search scheme \cite{boneh2004public} is adopted to secure the search of EMRs and to ensure identity privacy.

\textbf{Public blockchain-based eHealth systems (\cite{azaria2016medrec}, \cite{yang2017blockchain}, \cite{zhang2016secure}, \cite{cao2019cloud} and \cite{chen2019blockchain}).} In order to prompt patients to engage in the details of their healthcare and restore agency over their medical data, Azaria et al. \cite{azaria2016medrec} proposed a decentralized record management system to handle EMRs. The system utilized smart contracts to manage medical records of patients. With modular design, patients can make access control rules and share their EMRs with different institutions. The authors also designed reward system to motivate researchers and public health authorities to participate in the network as blockchain miners. 

Based on MedRec, Yang and Yang \cite{yang2017blockchain} used signcryption and attribute-based authentication to enable the secure sharing of healthcare data. EMRs are  encrypted with a symmetric key, which further encrypted with an attribute key set. The concatenation of both ciphertexts (encrypted EMRs and encrypted key) is signed with a private key. For data accessing, a user verifies the signature and performs EMRs decryption to get the plaintext EMRs.

Cao et al. \cite{cao2019cloud} proposed a secure cloud-assisted eHealth system to protect outsourced EMRs from illegal modifications. The key idea of the system is that the EMRs can only be outsourced by authenticated participants and each operation on outsourced EMRs is integrated into the public blockchain as a transaction. The system took into account the situations of a single doctor and multiple doctors and utilized key exchange protocol to protect the privacy of EMRs. The tamper-proofing property of blockchain guaranteed the correctness and integrity of EMRs.
\section{Preliminaries}
In this section, we review some basic preliminaries used in this paper, including cryptographic hash functions, chameleon hash functions, proxy re-encryption schemes and the Proof-of-Reputation blockchain system.
\subsection{Hash function and chameleon hash function}
\textbf{Cryptographic Hash Function.} A cryptographic hash function $H(\cdot)$ is a deterministic mathematical algorithm that maps an arbitrary-length string to a fixed-length bit string, i.e. $H(m)=h$, where $m$ is the message and $h$ is the hash value. In theoretical cryptography, the security of a cryptographic hash function has been defined using the following properties \cite{rogaway2004cryptographic}:
\begin{itemize}
	\item \textbf{Pre-image resistance.} Given a hash value $h$ it should be difficult to find any message $m$ such that $h=H(m)$.
	\item \textbf{Second pre-image resistance.} Given an input $m_1$, it should be difficult to find a different input $m_2$ such that $H(m_1)=H(m_2)$.
	\item \textbf{Collision resistance.} It should be difficult to find two different messages $m_1$ and $m_2$ such that $H(m_1)=H(m_2)$.
\end{itemize}

In some scenarios \cite{ateniese2017redactable}, the authorized entities should be allowed to easily find hash collisions to replace the wrong or harmful messages. To achieve this function Krawczyk and Rabin put forward the concept of chameleon hashing \cite{krawczyk1998chameleon} on the notion of chameleon commitments.

\textbf{Chameleon hash function.} Informally, a chameleon hash \cite{khalili2020efficient} is a cryptographic hash function that contains a trapdoor: Without the trapdoor, it should be hard to find collisions, but knowledge of the trapdoor information allows collisions to be generated efficiently. A chameleon hash function is a tuple of efficient algorithms $\mathcal{CH}=(\mathtt{HGen,Hash,HVer,HCol})$ specified as follows.
\begin{itemize}
	\item $(hk,tk)\leftarrow\mathtt{HGen}(1^\mathcal{K})$. The key generation algorithm $\mathtt{HGen}$ takes as input the security parameter $\lambda \in \mathbb{N}$, and outputs a public hash key $hk$ and a secret trapdoor key $tk$.
	\item $(h,R)\leftarrow \mathtt{Hash}(m,hk)$. The hashing algorithm $\mathtt{Hash}$ takes as input the hash key $hk$, a message $m\in \mathcal{M}$, and returns a pair $(h,R)$, where $r\in \mathcal{R}_{hash}$ denotes the implicit random coin used to generate the hash value.
	\item $d=\mathtt{HVer}(hk,m,(h,R))$. The verification algorithm $\mathtt{HVer}$ takes as input a message $m\in \mathcal{M}$ and a pair $(h,R)$, and returns a bit $d$ that equals ``1" if $\mathtt{HVer}(hk,m,(h,R))=h$ (otherwise $d$ equals ``0").
	\item $R'\leftarrow\mathtt{HCol}(tk,(h,m,R),m')$. The collision finding algorithm $\mathtt{HCol}$ takes as input the trapdoor key $tk$, a valid tuple $(h,m,R)$, and a new message $m'\in \mathcal{M}$, and returns a new random coin $R'$ such that $\mathtt{HVer}(hk,m,(h,R))=\mathtt{HVer}(hk,m',(h,R'))=1$. If $(h,R)$ is not a valid hash for message $m$, then the algorithm returns $\perp$.
\end{itemize}
\subsection{Proxy re-encryption}
In order to achieve privacy-preserving data sharing, we introduce the proxy re-encryption scheme. In a proxy re-encryption (PRE) scheme, the proxy converts the ciphertext encrypted with the delegator's public key into a ciphertext that can be decrypted with the delegatee's secret key through a re-encryption key from the delegator. A PRE scheme is illustrated as follows.
\begin{itemize}
	\item $(par)\leftarrow \mathtt{Setup}(1^\mathcal{K})$. Accepting an input security parameter $\mathcal{K}$, public parameters $par$ is produced by this algorithm.
	\item $(Rpk_i,Rsk_i)\leftarrow\mathtt{KeyGen}(par)$. This algorithm accepts the public parameter $par$ as input, and produces a pair of public key/private key $(Rpk_i,Rsk_i)$.
	\item $rk\leftarrow\mathtt{ReKeyGen}(par, Rpk_j,Rsk_i)$. In this algorithm, user $i$ takes a private key $Rsk_i$, the public parameter $par$ and the user $j$'s public key $Rpk_j$ as input, and produces $rk$.
	\item $C_i\leftarrow\mathtt{Enc}(par,Rpk_i,m)$. On input a user's $Rpk_i$, a message $m$ and public parameter $par$, this algorithm returns the original ciphertext$C_i$.
	\item $C_{i\rightarrow j}\leftarrow\mathtt{ReEnc}(rk)$. This algorithm is operated by a semi-honest proxy. On input the original ciphertext $C_i$ and a re-encryption key $rk$, this algorithm outputs the ciphertext $C_{i\leftarrow j}$ which can be decrypted by user $j$.
	\item $m\leftarrow\mathtt{Dec}_1(C_i,Rsk_i)$. This algorithm takes as input the ciphertext $C_i$ and the private key $Rsk_i$ of user $i$, and outputs the plaintext $m$.
	\item $m\leftarrow\mathtt{Dec}_2(C_{i\rightarrow j},Rsk_j)$. This algorithm inputs the ciphertext $C_{i\rightarrow j}$ and user $j$'s private key $Rsk_j$, and outputs the plaintext $m$.
\end{itemize}
\subsection{Blockchain and Proof-of-Reputation blockchain system}
The blockchain technology is a data structure and serves as a distributed ledger in which multiple transactions are maintained by trustless nodes in a P2P network. Information may include data records of different types, such as cryptocurrency transactions, smart contracts and account balances. The most successful blockchain system is the Proof-of-Work (PoW) blockchain underlying Bitcoin, where miners solve crypto-puzzles via hash computation. However, the PoW blockchain is resource-consuming and vulnerable to bribery attacks or flash attacks which allow attackers quickly gain computational power to control the blockchain network. To mitigate these attacks, Yu et al \cite{yu2019repucoin} proposed a Proof-of-Reputation (PoR) blockchain, RepuCoin, which considers miners' integrated power rather than using instantaneous mining power to select members. The integrated power is calculated using the total amount of valid work contributed by miners to the system. We make use of notations in \cite{garay2015bitcoin} to describe the basics of a PoR blockchain.
\begin{figure}[h]
	\centering
	\includegraphics[height=4cm,width=12cm]{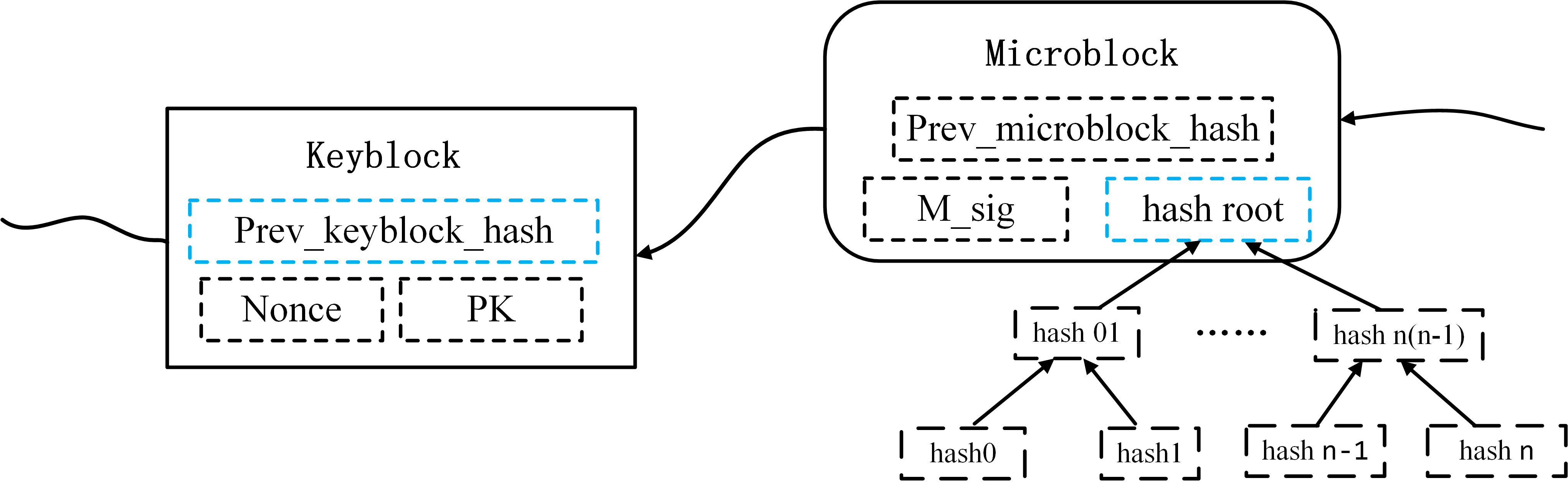}
	\caption{The block structure of \cite{yu2019repucoin}.}
	\label{fig1}
\end{figure}

In a blockchain system, a transaction is the basic unit of transferring accounts from a sender to a receiver. It is defined as the form of $tx=<address_{input},\\address_{output},data,sig>$, where $address_{input}$ indicates the sender's address, $address_{output}$ is the receiver's address, $data$ is the optional fields that can be padded with some extra data and $sig$ represents the sender's signature.  There are two kinds of blocks in a PoR blockchain \cite{yu2019repucoin}, namely keyblocks and microblocks. A miner creates keyblocks by solving the Bitcoin mining puzzle and may become a leader for a period of time. A leader can then verify and include transactions into microblocks directly. We call the microblocks are bonded with the keyblocks. As shown in Fig. \ref{fig1}, a keyblock is a triple of the form $KB=<s,nonce,pk>$ while a microblock is the form of $MB=<prevhash,tx,sig>$. Here $s\in \{0,1\}^\mathcal{K}$, $nonce \in\mathbb{N}$, $pk$ is the blockchain public key of the current leader, $prevhash$ is the hash value of the previous microblock, $tx$s denote transactions and $sig$ is the signature signed by using $pk$. Block $KB$ and $MB$ are valid if
\begin{itemize}
	\item $\mathtt{validBlock}^D(KB)=H(nonce,G(s,pk))<D$.
	\item $sig$ is the signature signed by using $pk$.
	\item Transactions contained in $MB$ are valid.
\end{itemize} 

\begin{figure}
	\centering
	\includegraphics[height=1.7cm,width=9.5cm]{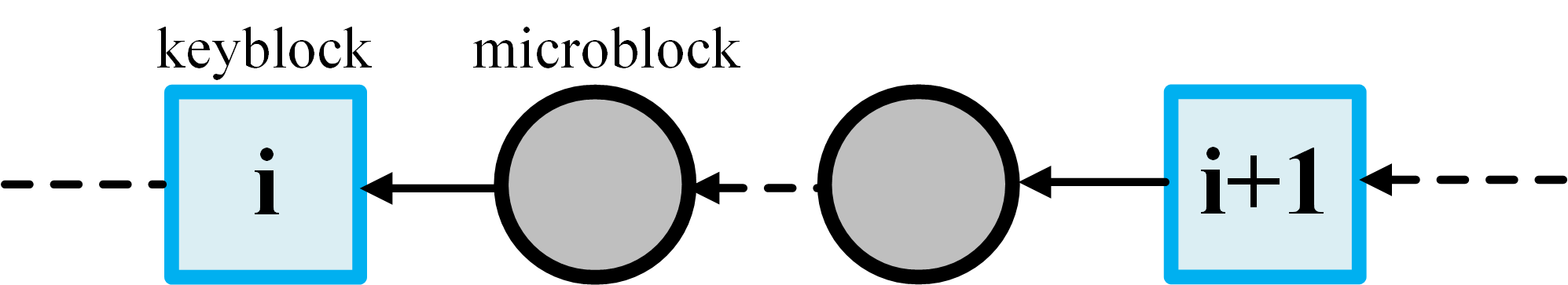}
	\caption{The chain structure in PoR blockchain \cite{yu2019repucoin}.}
	\label{fig2}
\end{figure}

Here $H(\cdot)$ and $G(\cdot)$ are cryptographic hash functions with output in $\{0,1\}^{\mathcal{K}}$, and the parameter $D\in\mathbb{N}$ is the block's difficulty level.  As illustrated in Fig. \ref{fig2}, we call a sequence of blocks a blockchain or simply a chain $\mathcal{C}$. 
The rightmost keyblock is the head of the chain, denoted $head(\mathcal{C})$. Any chain $\mathcal{C}$ with a head $\mathtt{Head}(\mathcal{C})=<s,nonce,pk>$ can be extended to a new longer chain $\mathcal{C}'$ by attaching a valid keyblock $KB'$ or microblock $MB'$ such that $\mathcal{C}'=\mathcal{C}||KB'$ or $\mathcal{C}'=\mathcal{C}||MB'$. The function $\mathtt{len}(\mathcal{C})$ denotes the length of a chain $\mathcal{C}$. Note that microblocks do not contain any proof of work. 

We define the consensus mechanism of a PoW blockchain system as the form $\mathcal{CM}=<mining,policy>$. In the $mining$ process, miners try to calculate a hash value $H(nonce,G(s,tx))$ less than $D$ to create a new block. In order to ensure the consistency of a PoW blockchain, all miners accept the $policy$ that the longest chain is the only chain. While in a PoR blockchain system, the consensus mechanism is defined as $\mathcal{CM}=<mining,BFT(R)>$. That is, in $mining$ process, miners calculate $H'(nonce,G(s,pk))$ less than $D$ to accumulate their reputation scores. Then the miners with the top $n$ reputation scores execute the Byzantine fault tolerance (BFT) protocol to generate new blocks.
\section{System and threat model}
In this section, we first delineate the workflow of SPChain. Thereafter we present the adversary model and the security requirements respectively.  
\subsection{High-level overview}
\begin{figure}[htbp]
	\centering
	\includegraphics[height=5cm,width=9cm]{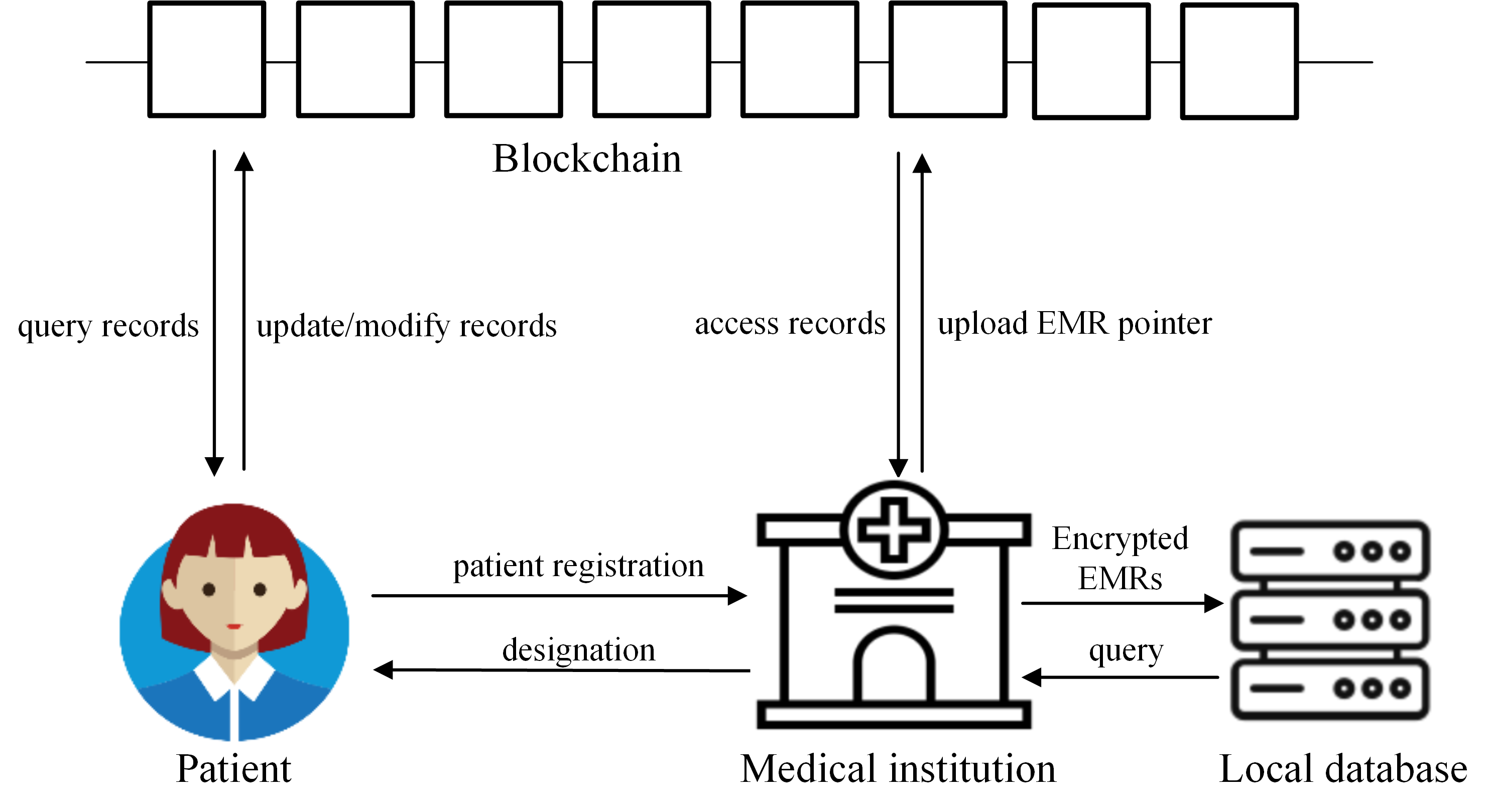}
	\caption{The architecture of SPChain.}
	\label{fig3}
\end{figure}

As shown in Fig. \ref{fig3}, there are two participants in our system: patients and medical institutions. Patients and medical institutions possess their own blockchain addresses of the form $<(pk,sk),address>$, where $(pk,sk)$ is a blockchain key pair and $address$ is the entity's blockchain address. Each entity also holds a PRE key pair $(Rpk,Rsk)$ to share EMR in a privacy-preserving manner. The procedure that patients consult medical institutions in SPChain is illustrated as follows.

Firstly, a patient $\tilde{i}$ sends a transaction to register into medical institution $\tilde{M}$. Upon receiving the register transaction, $\tilde{M}$ generates EMRs for $\tilde{i}$. After being encrypted with the patient $\tilde{i}$'s PRE public-key $Rpk_{\tilde{i}}$, the EMRs are stored in the medical institution's local database. The hash values and pointers of the encrypted EMRs are then uploaded to the blockchain via transactions. Then $\tilde{M}$ participates in mining to commit transactions into blocks. In return, the winning miners gain rewards and reputation scores which can be used to apply for the patients' EMRs.

When the patient $\tilde{i}$ wants to share EMRs to medical institution $\tilde{N}$, $\tilde{i}$ should generate re-encrypted key $rk$ by using own private key and the public key of $\tilde{N}$. Then the decryption right of the ciphertext can be delegated to $\tilde{N}$. In the case of misdiagnosis, patients send transactions to label the wrong records. Also, $\tilde{i}$ can access EMR history in SPChain, and do not need double registration. 
\subsection{Adversary model}
In SPChain, medical institutions work as miners to mine new blocks to accumulate their reputation scores. That is, the medical institutions first solve a cryptographic puzzle to join the consensus group. Then the medical institutions with the top $n$ reputation scores of the consensus group execute the Byzantine fault tolerance (BFT) protocol to generate new blocks. Therefore, we consider an adversary (a.k.a. Byzantine) who can drop, delay, re-order, insert or modify messages arbitrarily. The medical institutions can also collude with others to model a malicious real organization capable of deploying a significant number of virtual miners under its direct dependence.

In consequence, the consensus group can be infiltrated by adversaries. However, we hold the assumption that the number of Byzantine nodes in the system does not exceed $\frac{1}{3}$ the total number of group members whose collective reputation is less than $\frac{1}{3}$ of the cumulative reputation of the consensus members, which is the same as in RepuCoin \cite{yu2019repucoin}.

\subsection{Design goals}
Under the adversary model mentioned above, SPChain should satisfy the following security requirements.

\textbf{Confidentiality.} SPChain achieves EMRs sharing for patients without revealing sensitive data of patients. That is, the contents of EMRs should not be recovered by unauthorized medical institutions or attackers.

\textbf{Patient centric sharing.} In SPChain, patients take control of their EMRs. The unauthorized medical institutions cannot access the patients' EMRs.

\textbf{Quick retrieval.} SPChain provides special storage structure for patients such that patients can retrieve their medical history quickly.

\textbf{Label and correctness.} SPChain provides special transactions for authorized medical institutions to label wrong EMRs. Also, SPChain guarantees the integrity of EMRs, which means that any illegal modifications on EMRs should be detected by the system. 

\textbf{Resistance to blockchain attacks.} SPChain should prevent attackers from launching blockchain attacks, such as 51\% attacks under the proposed adversary model.
\begin{itemize}
\item \textit{51\% attacks \cite{apostolaki2017hijacking} and flash attacks \cite{bonneau2016buy}}. An attacker can obtain a temporary majority of computing power by renting enough mining capacity, which would break the security assumption of Proof-of-Work based blockchain systems. 

\item \textit{Selfish mining attacks (block withholding attack) \cite{sapirshtein2016optimal}}. In this case, an attacker controls a significant amount $(>25\%)$ of mining power in the system. Instead of publishing the mined block to the network, the selfish miner mines the block continuously maintaining its track. The attacker only publishes the chain of the transaction to increase the amount of revenue earned.
\item \textit{Sybil attacks.} In permissioned blockchains that without proof of work, an attacker can launch Sybil attacks to create multiple identities for voting and thus gain an advantage in consensus.
\end{itemize}
\textbf{Resistance to SPChain attacks.} SPChain should prevent attackers from launching the following SPChain attacks under the proposed adversary model.
\begin{itemize}
\item \textit{Reputation fraud attacks}. A malicious medical institution may create "zombie" patient nodes to send fake transactions to increase its reputation scores. 
\item \textit{Inhibition attacks}. When a medical institution of the consensus group becomes a leader successfully, it may only package its own transactions and ignore other medical institutions' transactions on purpose to increase the reputation scores. 
 \end{itemize}                                 
\section{Construction of SPChain}
In this section, we present details describing the different concepts and modules underlying SPChain. We first introduce the basic elements of SPChain. Thereafter, we describe how to integrate the SPChain with eHealth systems to achieve patient centric medical data sharing in a privacy-preserving manner.
\subsection{Basic elements in SPChain}
\begin{figure}[htbp]
	\centering
	\includegraphics[height=4cm,width=11cm]{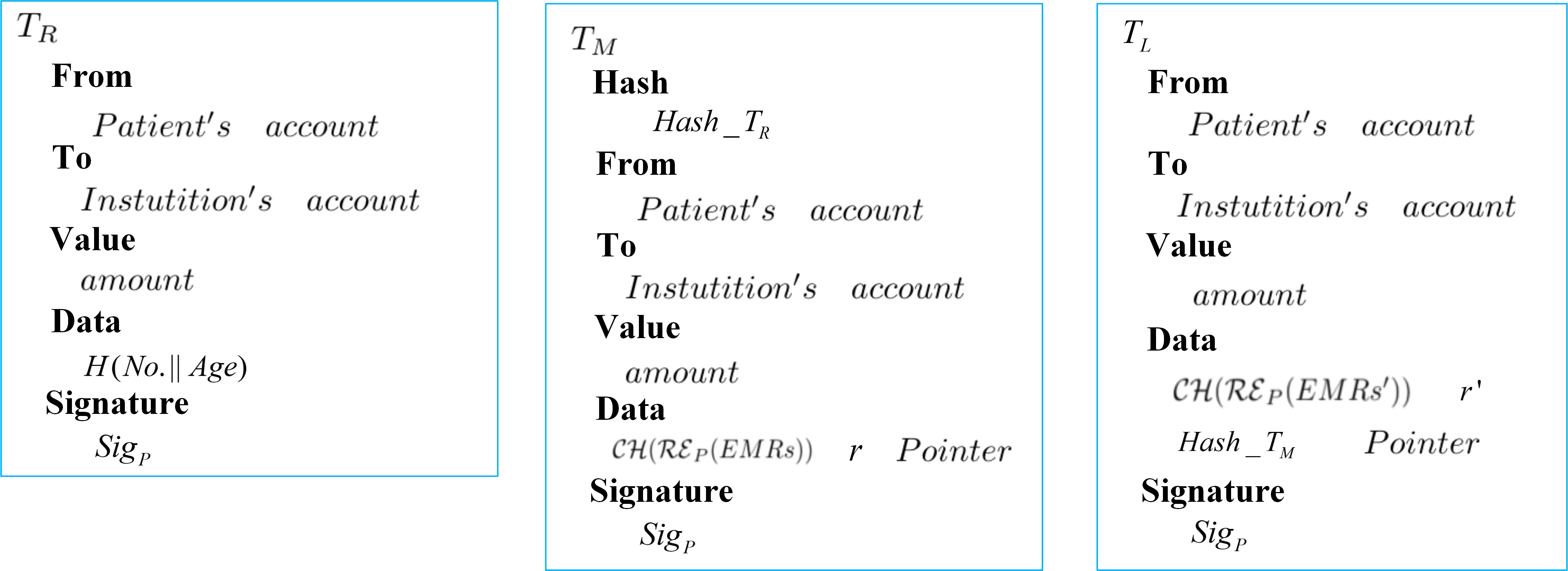}
	\caption{The structure of transactions in SPChain.}
	\label{fig4}
\end{figure}

\textbf{Transaction.} As shown in Fig. \ref{fig4}, there are three types of transactions in our system, namely register transactions, medical transactions and label transactions. Transactions in SPChain are presented as the form of $(Type,address_{input},\\ address_{output},data,sig)$, where $Type$ denotes the type of a transaction, $address_{input}$ and $address_{output}$ represent the blockchain address of the sender and the receiver respectively, $data$ identifies the contents in different types of transactions and $sig$ specifies the signature of the transaction sender. In our system, register transactions are packed into keyblocks while medical transactions and label transactions are attached to the microblocks bound to the patients. We use $\mathcal{RE}$ to represent the encrypted ciphertext in a PRE scheme. 
\begin{itemize}
\item \textit {Register transaction.} This transaction is send to the medical institution which the patient wants to be treated at the first time. We denote a register transaction as $T_R=(register,H(No.||Age),sig_P)$, where $register$ specifies the transaction type, $H(No.||Age)$ denotes the hash value of a patient's medical record number, age and other auxiliary information, and $sig_P$ is the signature of the patient. Every patient should send this transaction to register in SPChain.   
\item\textit {Medical transaction.} This transaction is send by patients to upload records to the blockchain. We define a medical transaction $T_M=(medical,\mathcal{CH}\\({\mathcal{RE}_P}(EMRs))||r||pointer,sig_P)$, where $medical$ indicates that this is a medical transaction, $\mathcal{CH}({\mathcal{RE}_P}(EMRs))$ represents the chameleon hash value of the encrypted EMRs, $pointer$ is the pointer to the encrypted EMRs stored in the local database and $sig_P$ is the signature of the patient.   
\item\textit {Label transaction.} When medical errors occurs, patients send this transaction to label the wrong records. We define a label transaction $T_M=(label,\\H(T_M),\mathcal{CH}({\mathcal{RE}_P}(EMRs'))||r'||pointer,sig_P)$, where $label$ indicates the transaction is a label transaction, $H(T_M)$ is the transaction hash value of the medical transaction to be labeled, $\mathcal{CH}({\mathcal{RE}_P}(EMRs'))$, $r'$, $pointer$ and $sig_P$ are the same as the above definitions.
\end{itemize}

\begin{figure}[htbp]
	\centering
	\includegraphics[height=5cm,width=12cm]{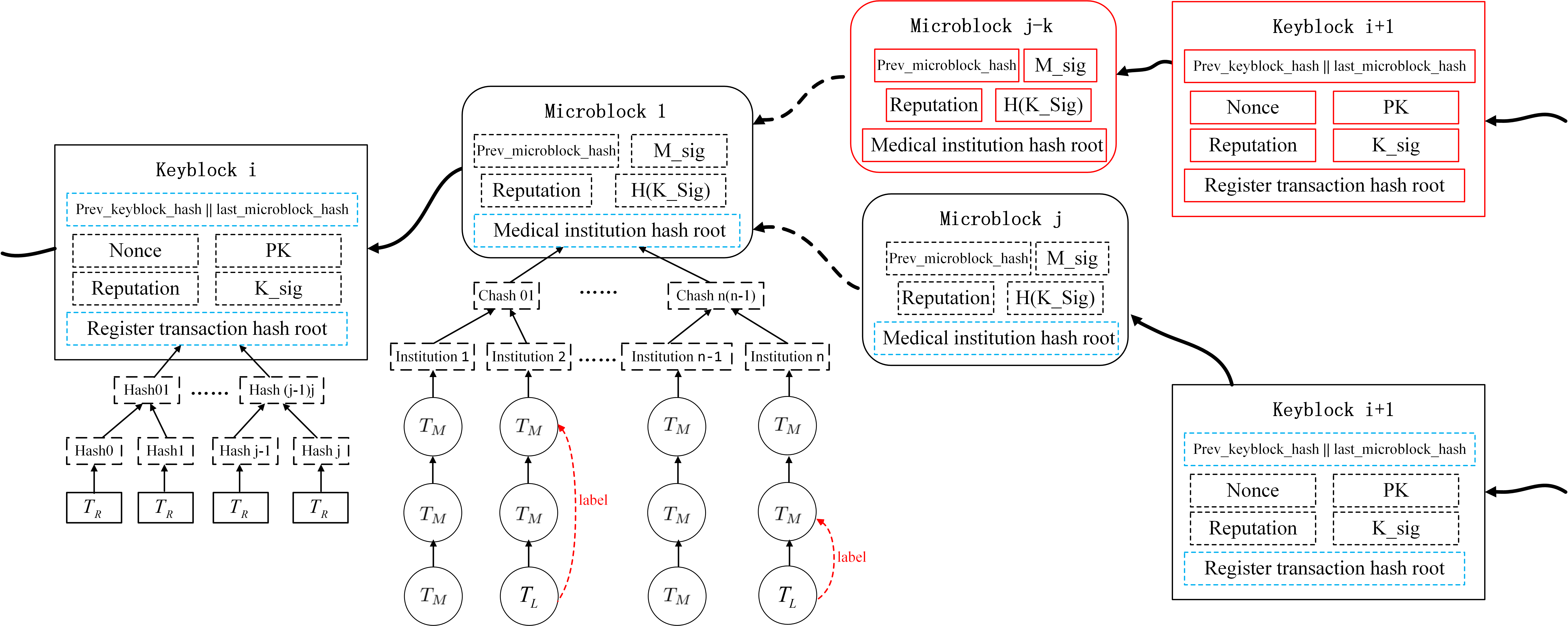}
	\caption{The structure of blocks in SPChain.The red blocks is a fork of the main blockchain (red blocks).}
	\label{fig5}
\end{figure}

{\bfseries Blocks.} As shown in Fig. \ref{fig5}, SPChain contains two different blocks, namely keyblocks and microblocks. In RepuCoin \cite{yu2019repucoin} and Bitcoin-ng\cite {eyal2016bitcoin}, the keyblocks do not contain transactions, while the keyblocks of SPChain contain register transactions $T_R$ such that $KB=<s,T_R,nonce,pk>$. The {\it{Register transaction hash root}} in a keyblock is calculated by a cryptographic hash function such as SHA-256 used in Bitcoin, which means that the registers transactions contained in the keyblock cannot be modified without changing the hash root.

\begin{figure}[htbp]
	\centering
	\includegraphics[height=4cm,width=10cm]{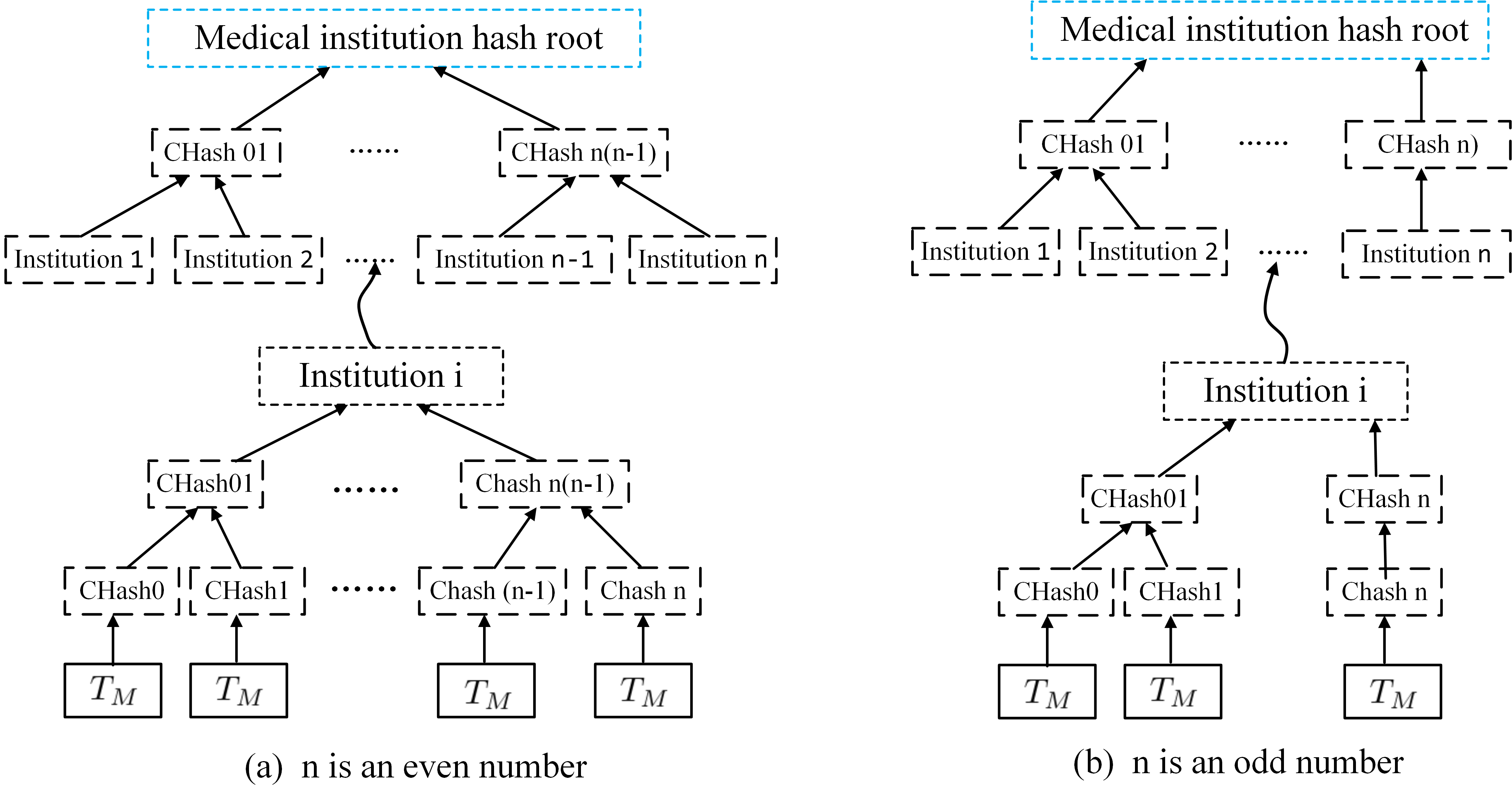}
	\caption{The calculation of the {\it{Medical institution hash root}} in a Microblock.}
	\label{fig6}
\end{figure}

We define a microblock $MB=<prevhash,T_M,T_L,sig>$ as a patient block which contains all EMRs of the patient in different medical institutions. In order to facilitate the retrieval of records, we use merkle tree structure to construct the institution hash root. We leverage the chameleon hash function to calculate the {\it{Medical institution hash root}} to update and retrieve EMRs of a patient in the microblock. The leaf nodes of the tree are the basic information (for example, the public key which is certified by authority) of medical institutions. Fig. \ref{fig6} details the calculation of hash root. There are two cases in the calculation, an even number $n$ and an odd number $n$. Medical transactions and label transactions are attached behind the basic information in chronological order.

\begin{figure}[htbp]
	\centering
	\includegraphics[height=5.5cm,width=10cm]{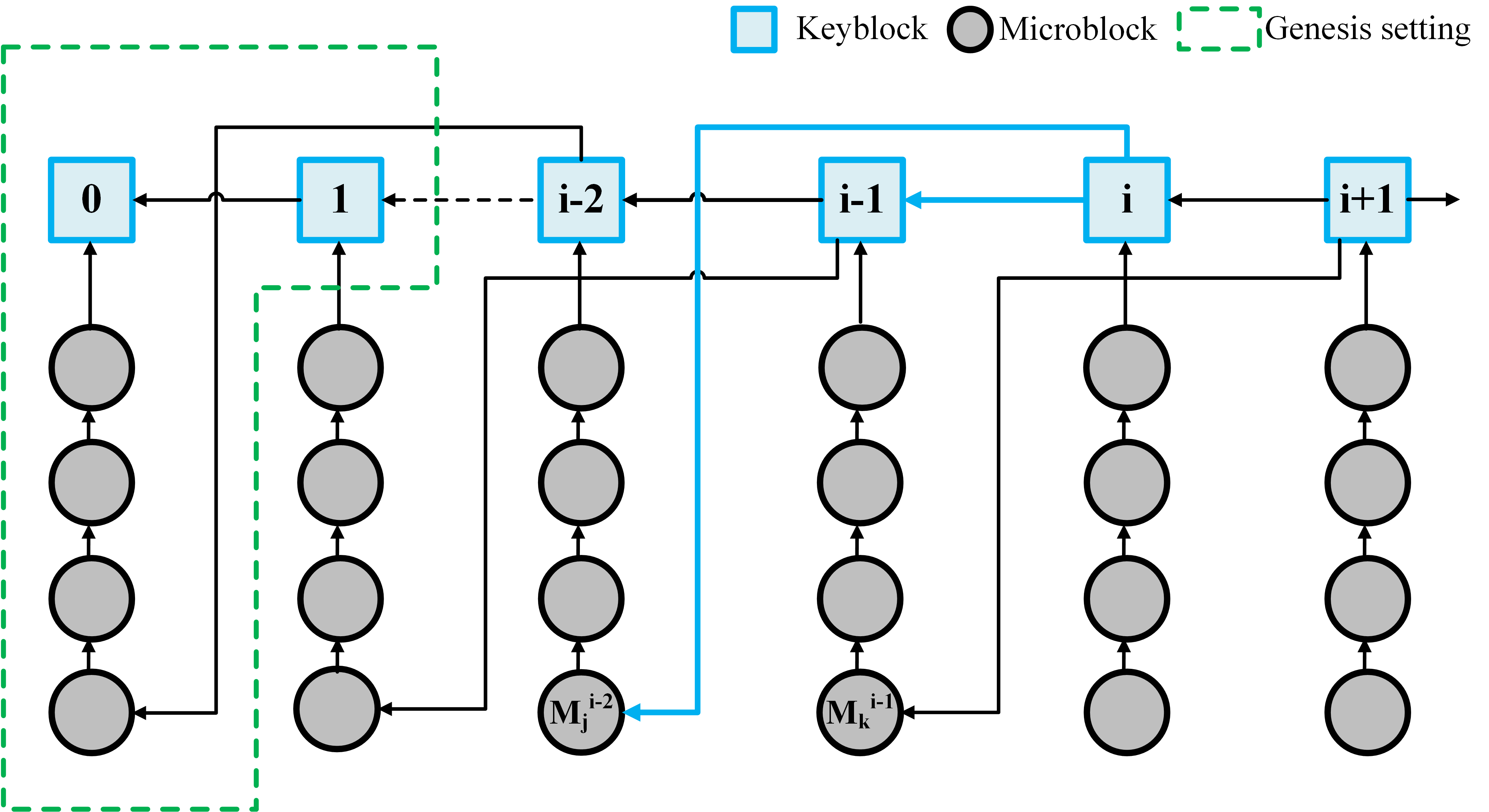}
	\caption{The chain structure in SPChain. From the blue arrows we can see that the input of the keyblock $KB_i$ is coming from two parts, the hash value of the microblock $MB^{i-2}_j$ and the hash value of the keyblock $KB_{i-1}$. The genesis part is set by the system management.}
	\label{fig7}
\end{figure}

{\bfseries Chain structure.} We regard the register transactions stored in the keyblock as the indexes of the bonded microblocks. That is, the microblocks that follow each keyblock correspond one-to-one with the register transactions in the keyblock. In this case, the chain structures in \cite{yu2019repucoin}, \cite{yu2021novel} and \cite{eyal2016bitcoin} do not suit for SPChain because the chain structure may fork as shown in Fig. \ref{fig5}, resulting in a mismatch problem between register transactions of the keyblock and the bonded microblocks. In order to avoid the mismatch, we devise a new chain structure which is illustrated in Fig. \ref{fig7}. Suppose a miner wants to mine the keyblock $KB_i$ to extend the chain $\mathcal{C}$ with $\mathtt{Head}(\mathcal{C})=KB_{i-1}$. The miner takes the hash values of microblock $MB_j^{i-2}$ and keyblock $KB_{i-1}$ as inputs and tries to calculate a hash value $H(nonce,G(MB_j^{i-2}),G(KB_{i-1}),pk)$ that less than $D$, where $H(\cdot)$, $G(\cdot)$ are cryptographic hash functions and $D$ is the block's difficulty level. This design avoids the mismatch problem because it gives the leader of keyblock $KB_{i-1}$ enough time to generate the bonded microblocks before the keyblock $KB_i$ is minded.
\begin {table}[h]
\centering\caption{The notations of reputation calculating.}
\label{tab1}
\small
\begin{tabular}{c|p{10cm}}
	\hline
	Symbol &            Description  \\
	\hline
	$L$      &the length of the current blockchain;\\
	\hline
	$c$    &the size of a block chunk, i.e. the number of keyblocks contained in a chunk, pre-defined by the system;\\
	\hline
	$l$  &$l=\lceil{\frac{L}{c}}\rceil$ is the number of keyblocks contained in a blockchain with length $L$;\\
	\hline
	$N$                        & the total number of the current microblocks;  \\
	\hline
	$T$                        & total transactions in blockchain;                    \\
	\hline
	$TML_i$ & the number of medical transactions and label transactions whose receiptor is miner in chunk $i$;  \\
	\hline
	$TR_i$    &  the number of register transactions whose receiptor is miner in chunk $i$;  \\  
	\hline
	$H$ & a binary presenting whether the miner is honest ("1") or not ("0");\\
	\hline               		
	$mean_i$     &the mean value of medical transactions and label transactions (if $i = TML$) or register transactions (if $i = TR$) created by a miner or a leader across all epochs in the blockchain, respectively;                                    \\
	\hline
	$s_i$             & the standard deviation corresponding to $mean_i$, for $i \in \{TML,TR\}$;  \\
	\hline
	$R_1$&   reputation score defined in RepuCoin;\\
	\hline
	$(a,\lambda)$           &reputation system parameters. \\
	\hline
\end{tabular}
\end{table}

\textbf{Consensus mechanism.} We define the consensus mechanism of SPChain as the form $\mathcal{CM}=<PoW,BFT(R)>$. That is, medical institutions try to calculate $H(nonce,G(s,pk))$ less than $D$ to accumulate their reputation scores $R$. Then the miners with the top $n$ reputation scores execute the Byzantine fault tolerance (BFT) protocol to generate new blocks. RepuCoin \cite{yu2019repucoin} gives a method to calculate the reputation score $R_1$ by evaluating the frequency of miners creating keyblocks and microblocks. In SPChain, we leverage the number of registered patients and medical records in different medical institutions to calculate the reputation score $R_2$. We call $R_1$ the blockchain reputation score and $R_2$ the medical reputation score. The reputation score of a medical institution is defined as $R=\frac{1}{2}(R_1+R_2)$ and can be regarded as the metric to assess whether a medical institution is worthy of obtaining patients' EMRs. The notions are defined in Table \ref{tab1} and $R_2$ is calculated in Algorithm \ref{alg:1}.
 \begin{algorithm}[h]
	\renewcommand{\algorithmicrequire}{\textbf{Input:}}
	\renewcommand{\algorithmicensure}{\textbf{Output:}}
	\caption{\hfill{The reputation algorithm}}
	\label{alg:1}
	\begin{algorithmic}[1]		
		\REQUIRE $L,c,l,TR_i,TML_i,R_1,a$ and $\lambda$
		\ENSURE The miners' reputation $R_2\in [0,1]$.
		\FOR {All medical institutions} 
		\STATE calculate $mean_{TR}=\frac{\sum_{i=1}^{l}TR_i}{N}$ 
		\STATE calculate $mean_{TML}=\frac{\sum_{i=1}^{l}TML_i}{T}$
		\STATE calculate $s_{TR}=\sqrt{\frac{1}{l}\cdot\sum_{i=1}^l(\frac{TR}{c}-\frac{\sum_{i=1}^{l}TR_i}{N})^2}$
		\STATE calculate $s_{TML}=\sqrt{\frac{1}{l}\cdot\sum_{i=1}^l(\frac{TML}{c}-\frac{\sum_{i=1}^{l}TML_i}{T})^2}$
		\ENDFOR
		\STATE Set $q_1=\frac{mean_{TR}}{1+s_{TR}}$ and  $q_2=\frac{mean_{TML}}{1+s_{TML}}$
		\STATE Define $x=q_1\cdot q_2\cdot L$
		\STATE Define $f(x)=\frac{1}{2}{(1+\frac{x-a}{{\lambda} +|x-a|})}$
		\STATE $R_2=min(1,H\cdot f(x))$
		\STATE $R=\frac{1}{2}(R_1+R_2)$
	\end{algorithmic}
\end{algorithm}

In order to prevent a malicious medical institution in the consensus group from launching inhibition attacks, we devise a transaction processing algorithm which stipulates that the transactions of medical institutions are verified proportionally according to their reputation. We illustrate the notations of reputation calculating in Table \ref{tab2} and detail the transaction processing algorithm to resist inhibition attacks in Algorithm \ref{alg:2}.

We call a round $l$ is a process where a keyblock and the corresponding microblocks are generated. In each round the keyblock is the sorting index of the following microblocks, which means the microblocks are mined in the order of the register transactions packed in the keyblocks. When sending medical or label transactions, patients should append rounds number to transactions to shard them in the consensus group.
\begin{figure}[h]
	\centering
	\includegraphics[height=2.5cm,width=8cm]{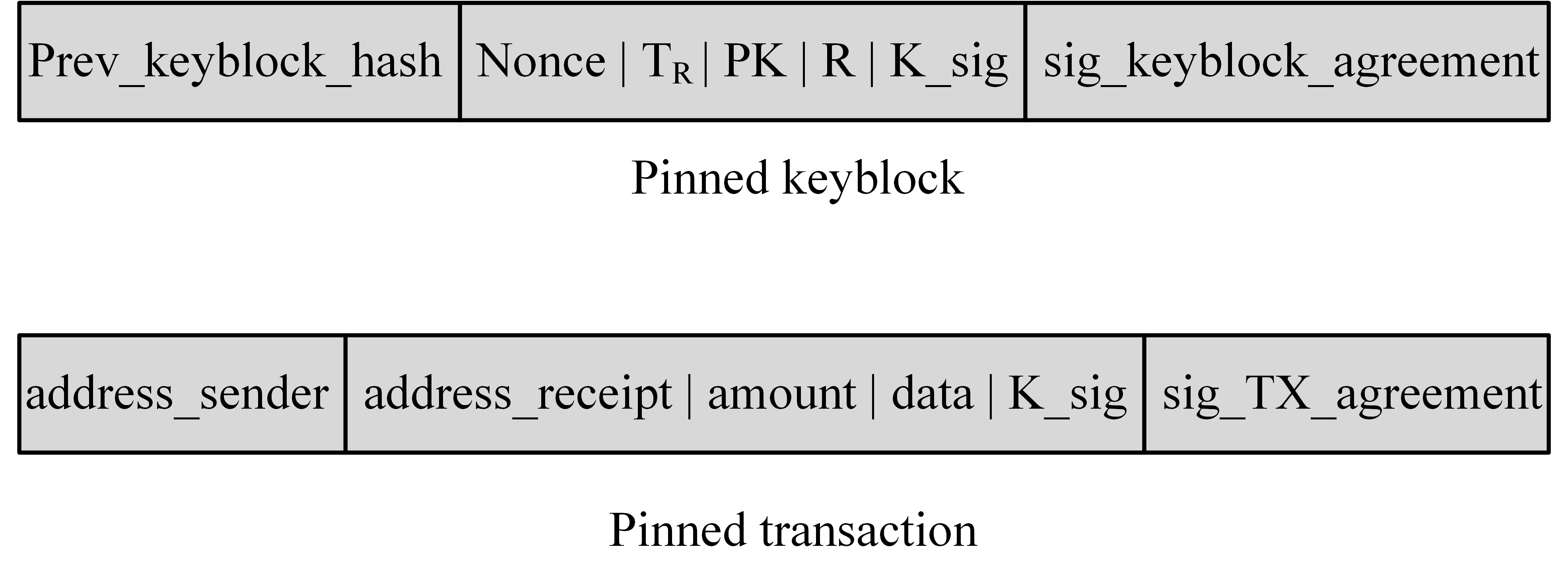}
	\caption{The formats of a pinned keyblock and a pinned transaction.}
	\label{fig8}
\end{figure}

To mitigate the fork problem, we use the pinned blocks mentioned in RepuCoin \cite{yu2019repucoin}. Fig. \ref{fig8} details the formats of a pinned keyblock and a pinned transaction. A pinned keyblock is a keyblock that is agreed upon and signed by the consensus group. A pinned keyblock is final and canonical, and all keyblocks that conflict with a pinned keyblock are considered invalid. Based on this definition, we also define the pinned transactions. Each time the transactions are generated by patients, the medical institutions collect the transactions sent to themselves, and propose them to the consensus group. The group verifies the received transactions and signs to the valid transaction. Then the medical institutions append the pinned transactions to the corresponding microblocks.

\textbf{Reward mechanism.} In SPChain the rewards can be divided into two parts, transaction fees and mining rewards. Medical institutions can define the determined amount of different type transactions to gain the corresponding transaction fees. After mining a pinned keyblock successfully, the medical institutions can get a reward contained the predefined mining rewards and the register transaction fees in the keyblock. Similar to keyblock rewards, microblock rewards also contain mining rewards and transaction fees, which are shared among the reputable medical institutions who create the microblocks and verify the contained transactions. 
\begin {table}[h]
\centering\caption{The notations of reputation calculating}
\label{tab2}
\small
\begin{tabular}{c|p{10cm}}
	\hline
	Symbol &            Description  \\
	\hline
	$m_i$      &the i-th medical institution; \\
	\hline
	$T_i$    &the transaction set of $m_i$;\\
	\hline
	$G$  &the consensus group;\\
	\hline
	$n$      & the total number of the medical institutions;  \\
	\hline
	$R_i$              & the reputation score of $m_i$;                    \\
	\hline
	$\Delta$ & the time interval;  \\
	\hline
	$k$    &  a nonce;  \\  
	\hline
	$T_m$ & the maximum number of transactions processed by consensus group at a time;\\
	\hline
\end{tabular}
\end{table}
\begin{algorithm}[htbp] 
\renewcommand{\algorithmicrequire}{\textbf{Input:}}
\renewcommand{\algorithmicensure}{\textbf{Output:}}
\caption{\quad\quad\quad  \quad \quad \quad \quad Transaction processing algorithm}
\label{alg:2}
\begin{algorithmic}[1]		
	\REQUIRE $m_i$, $R_i$, $T_i$, $k$, $p=\sum_{i=1}^{k}|T_i|$, $P=\sum_{i=1}^{k}T_i$, $\Delta$ and $G$. 
	\ENSURE The pinned transaction set $T$.
	\STATE $m_i$ sends $T_i$ within $\Delta$ to $G$  
	\STATE $G$ collects $T_i$ and forms a table $B=[T_1,\dots,T_n]$
	\STATE $G$ selects $t_i\in T_i$ according to reputation ranking , where $t_i= 10\lfloor R_i \rfloor$
	\STATE \textbf{Case 1:} $k\leq N$\\
	$G$ picks $T_i$ from $B$ until $p=\sum_{i=1}^{k}|T_i|$
	\STATE \textbf{Case 2:} $k>N$\\
	$G$ picks transactions from the beginning medical institution until $p=\sum_{i=1}^{k}|T_i|$\\	   
	\STATE $P$ is the final verified set $T$
\end{algorithmic}
\end{algorithm}
\subsection{Description of SPChain}
\begin{figure}[htbp]
	\centering
	\includegraphics[height=8cm,width=12cm]{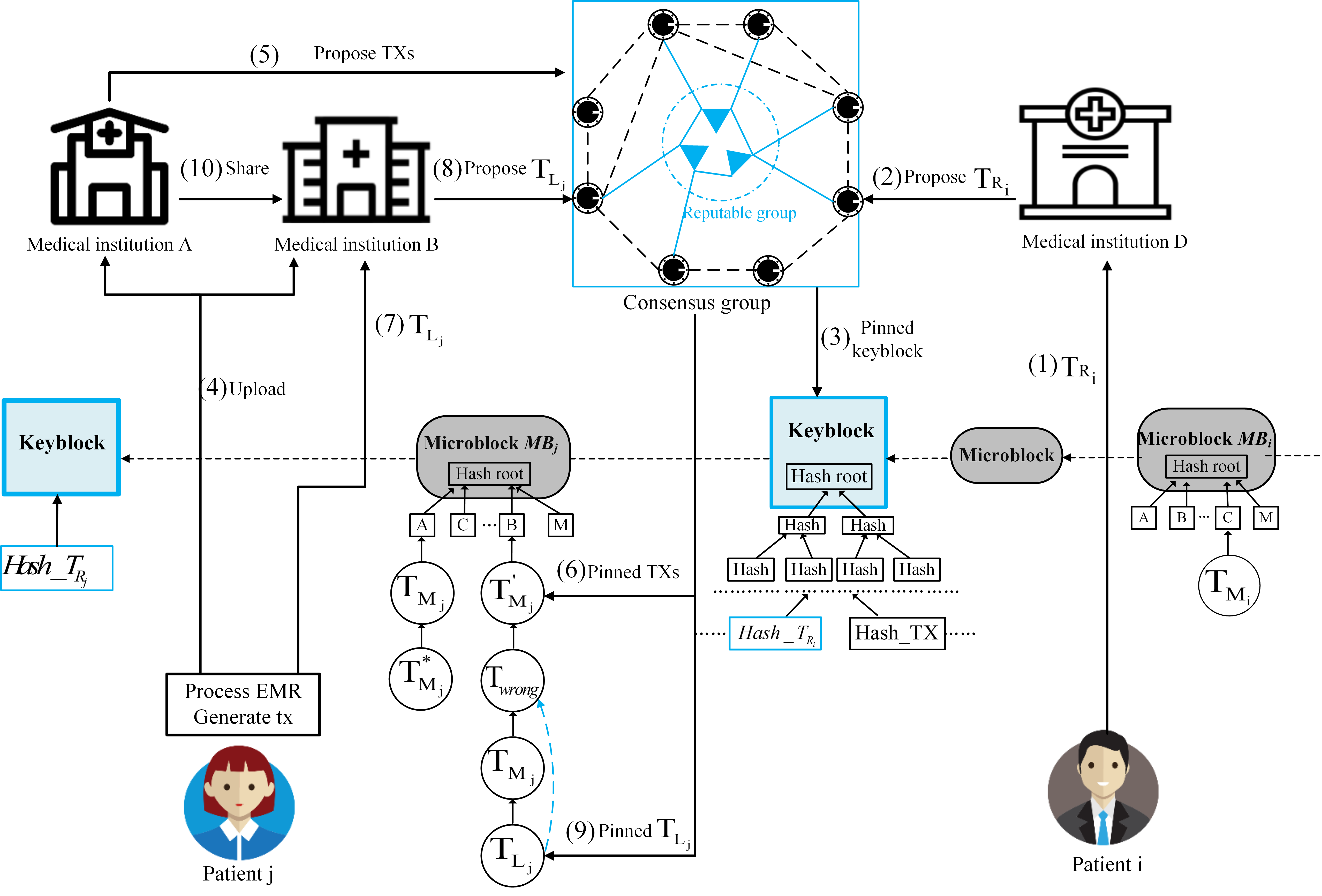}
	\caption{The orchestration of SPChain.}
	\label{fig9}
\end{figure}

In this part, we present the detailed workflow of SPChain upon the above basics. Every entity (e.g. patients and medical institutions) in SPChain holds a blockchain account of the form $<(pk,sk),address>$. Fig.\ref{fig9} details the orchestration of SPChain. The system is consist of the following phases, namely \textbf{Setup}, \textbf{Register}, \textbf{Upload}, \textbf{Label}, \textbf{Share} and \textbf{Retrieval}, which are illustrated as follows.

\textbf{Setup}. This phase initializes the system parameters and generates accounts for entities. Firstly entities invoke PRE.$\mathtt{KeyGen}$ and bitcoin accounts algorithms to generate the PRE key pair $(Rpk,Rsk)$ and the bitcoin accounts $<(pk,sk),address>$ respectively. Then medical institutions initialize the chameleon hash function parameters by invoking the algorithm $\mathcal{CH}.\mathtt{HGen}$.
\begin{itemize}
\item 	Patients:\\
$(Rpk_p,Rsk_p)\leftarrow \mathtt{KeyGen}$,\\
$<(pk_p,sk_p),address_{pk_p}> \leftarrow Bitcoin(rand)$.
\item Medical institutions:\\
$(Rpk_m,Rsk_m)\leftarrow \mathtt{KeyGen}(par)$,\\
$<(pk_m,sk_m),address_{pk_m}> \leftarrow Bitcoin(rand)$,\\
$(hk,tk)\leftarrow \mathtt{HGen}(1^{\mathcal{K}})$.
\end{itemize}

\textbf{Register.} In the register phase, patients send register transactions to the medical institutions to register in SPChain.

($1$) Patient $\tilde{i}$ sends register transaction $T_{R_{\tilde{i}}}$ to medical institution $\tilde{C}$. The transaction contains proper register fees to $\tilde{C}$.

($2$) Medical institutions (miners) collect register transactions and pack them into keyblocks. Then Medical institutions propose keyblocks to consensus group.

($3$) The consensus group verifies the validity of the keyblocks, and  runs Byzantine agreement protocol to decide which keyblock is the final pinned keyblock (if multiple conflicting keyblocks are proposed). Then the reputable miner is selected to commit microblocks according to the register transactions in the keyblock.

\begin{figure}[h]
	\centering
	\includegraphics[height=8.5cm,width=10cm]{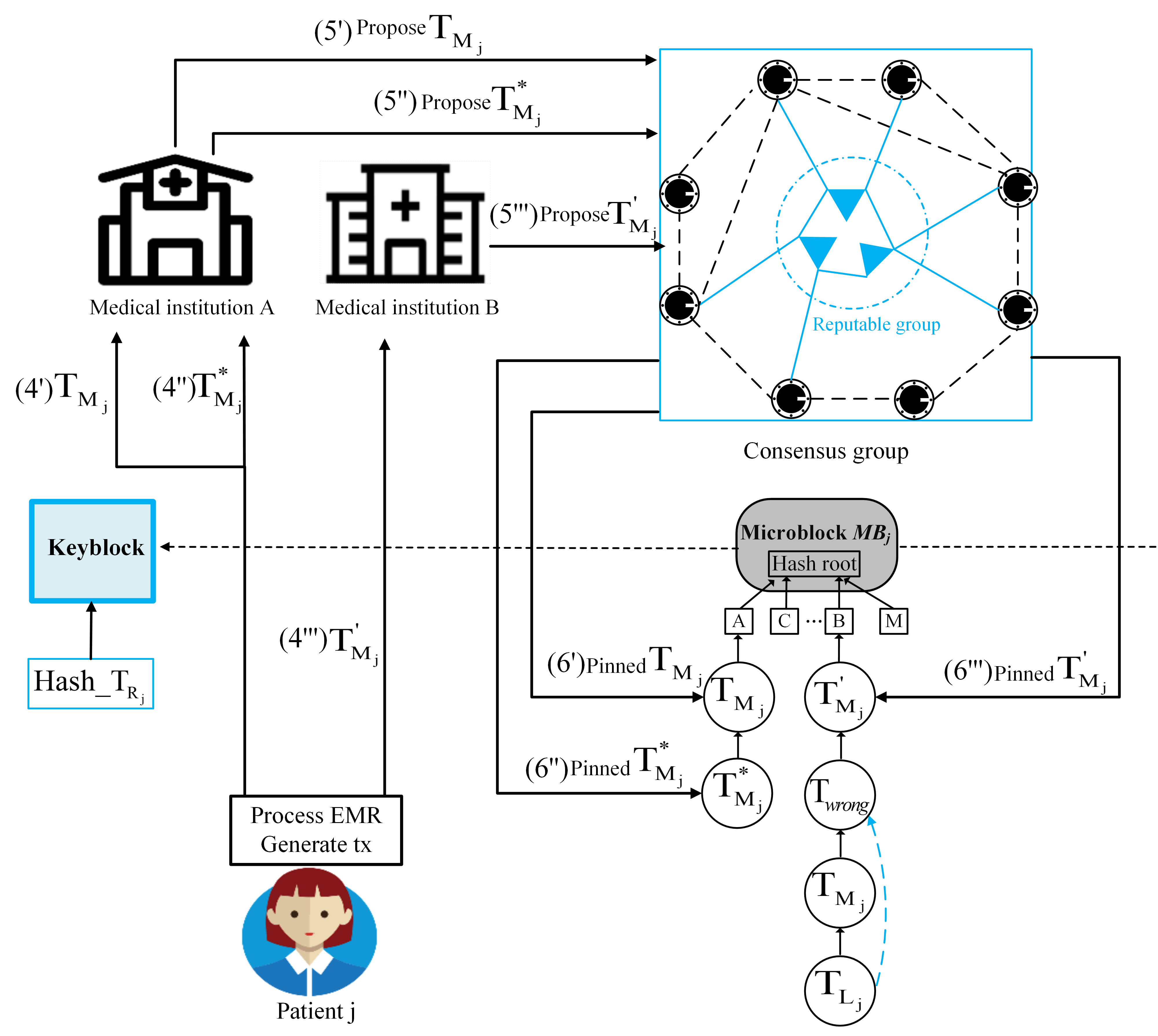}
	\caption{The three cases in $Upload$}
	\label{fig10}
\end{figure}

\textbf{Upload.} The serial number (4)-(6) given in Fig. \ref{fig9} illustrate the process of a patient uploading the EMRs to SPChain. There are three cases in this phase, patient $\tilde{j}$ is diagnosed in medical institution $\tilde{A}$ for the first time; patient $\tilde{j}$ updates EMRs in the same medical institution $\tilde{A}$; or patient $\tilde{j}$ is diagnosed in another department of medical institution $\tilde{B}$. We describe the three cases in detail in Fig.\ref{fig10}.

Case 1: Patient $\tilde{j}$ is diagnosed in medical institution $\tilde{A}$ for the first time.

($4'$) Patient $\tilde{j}$ registers in the system and is diagnosed in medical institution $\tilde{A}$. Firstly, $\tilde{A}$ generates EMR for $\tilde{j}$ and invokes PRE.$\mathtt{Enc}$ to encrypt the EMR with the PRE public key $Rpk_{\tilde{j}}$ of patient $\tilde{j}$. Then medical institution $\tilde{A}$ invokes $\mathcal{CH}.\mathtt{Hash}$ to generate chameleon hash value $(\mathcal{RE}_{Rpk_{\tilde{j}}}(EMR)||r)$ of the encrypted EMRs. Besides, $\tilde{A}$ generates a pointer that point to the local database, and sends them to $\tilde{j}$. 

Upon receiving $(\mathcal{RE}_{Rpk_{\tilde{j}}}(EMR)||r,pointer)$, patients $\tilde{j}$ generates transaction $T_{M_{\tilde{j}}}$ and sends it to the medical institution $\tilde{A}$.
Note that the ciphertext $(\mathcal{RE}_{Rpk_{\tilde{j}}}(EMR)$ is stored in the local database of medical institution $\tilde{A}$.

($5'$) The consensus group invokes the transaction processing algorithm (Algorithm \ref{alg:2}) to verify the transactions proportionally according to the medical institutions' reputation scores.

($6'$) The consensus group verifies the validity of the transactions and signs to them. To become pinned transactions, the transactions should not only get two-thirds of signatures, but also get more than two-thirds of the reputation scores. Then the medical institution $\tilde{A}$ appends the pinned transactions to the microblock $MB_{\tilde{j}}$.

Case 2: Patient $\tilde{j}$ updates EMRs in the same medical institution $\tilde{A}$. In this case, patient $\tilde{j}$ goes to the medical institution $\tilde{A}$ for treatment again and updates the EMRs on the basis of Case 1.

($4''$) The medical institution $\tilde{A}$ generates EMR* for $\tilde{j}$ and invokes PRE.$\mathtt{Enc}$ and $\mathcal{CH}.\mathtt{Hash}$ to get $\mathcal{RE}_{Rpk_{\tilde{j}}}(EMR^*)$, $\mathcal{CH}(\mathcal{RE}_{Rpk_{\tilde{j}}}(EMR^*))$ and $r^*$.

($5''$) Patient $\tilde{j}$ generates transaction $T_{M_{\tilde{j}}}^*$ and sends it to medical institution $\tilde{A}$. Then the consensus group collects the transactions, validates $T_{M_{\tilde{j}}}^*$ and signs it.

($6''$) After the transaction $T_{M_{\tilde{j}}}^*$ is pinned, medical institution $\tilde{A}$ appends the transaction to the microblock $MB_{\tilde{j}}$.

Case 3: Patient $\tilde{j}$ is diagnosed in other medical institution $\tilde{B}$. 

($4'''$) Medical institution $\tilde{B}$ generates EMR$'$ for $\tilde{j}$ and invokes PRE.$\mathtt{Enc}$ and $\mathcal{CH}.\mathtt{Hash}$ to obtain $\mathcal{RE}_{Rpk_{\tilde{j}}}(EMR')$, $\mathcal{CH}(\mathcal{RE}_{Rpk_{\tilde{j}}}(EMR'))$ and $r'$.

($5'''$) Then patient $\tilde{j}$ generates transaction $T_{M_{\tilde{j}}}'$ and sends it to medical institution $\tilde{B}$. The consensus group validates $T_{M_{\tilde{j}}}'$ and signs it.

($6'''$) After the transaction $T_{M_{\tilde{j}}}'$ is pinned, medical institution $\tilde{B}$ appends the transaction to the microblock $MB_{\tilde{j}}$.

\textbf{Label.} In this phase, patient $\tilde{j}$ labels the wrong EMRs in case of misdiagnosis. The transaction which contains the wrong EMR should be labeled by a label transaction.  
 
($7$) The medical institution $\tilde{B}$ generates the correct EMR$''$ for $\tilde{j}$ and invokes PRE.$\mathtt{Enc}$ and $\mathcal{CH}.\mathtt{Hash}$ to generate $\mathcal{RE}_{Rpk_{\tilde{j}}}(EMR'')$, $\mathcal{CH}(\mathcal{RE}_{Rpk_{\tilde{j}}}(EMR''))$ and $r''$. 

($8$) After that patient $\tilde{j}$ generates transaction $T_{L_{\tilde{j}}}$ and sends it to medical institution $\tilde{B}$. Then the consensus group validates $T_{L_{\tilde{j}}}$ and signs it.
 
($9$) Finally the pinned transaction $T_{L_{\tilde{j}}}$ labels the wrong transaction in microblock $MB_{\tilde{j}}$. And patients can verify the new transaction through the new proof $r''$.

\textbf{Share and retrieval.} Patient $\tilde{j}$ shares EMRs to the medical institution $\tilde{D}$ and retrieve EMRs as follows.

($10$) Suppose patient $\tilde{j}$ wants to go to medical institution $\tilde{D}$ for diagnosis. $\tilde{j}$ wants to share his/her EMRs in medical institution $\tilde{A}$ to medical institution $\tilde{D}$ to get a better treatment. Firstly, $\tilde{j}$ takes the public parameter $par$, his/her private key $Rsk_{\tilde{j}}$ and the medical institution $\tilde{D}$'s public key $Rpk_{\tilde{D}}$ as input and invokes PRE.$\mathtt{ReKeyGen}$ to generate the re-encryption key $rk_{\tilde{j}\rightarrow \tilde{D}}$. Then he/she transmits $rk_{\tilde{j}\rightarrow \tilde{D}}$ to the medical institution $\tilde{A}$ which is the first medical institution he/she registered. Using $rk_{\tilde{j}\rightarrow \tilde{D}}$ the medical institution $A$ invokes PRE.$\mathtt{ReEnc}$ to generate the ciphertext $C_{\tilde{j}\rightarrow \tilde{D}}$, and transmits the ciphertext to medical institution $\tilde{D}$. Then $\tilde{D}$ invokes PRE.$\mathtt{Dec}_2$ to obtain the EMRs of patient $\tilde{j}$. In addition, the medical institution $\tilde{D}$ can apply for EMRs from patients $\tilde{j}$. With a reputation score higher than the threshold set by $\tilde{j}$, $\tilde{D}$ can require access to $\tilde{j}$'s EMRs from medical institution $\tilde{A}$.
			
The patient $\tilde{j}$ can also access his/her EMR history. In this case $\tilde{j}$ submits his block number to any medical institutions and obtain the entire history of diagnosis.

\section{Performance evaluation and security analysis.}
In this section, we first evaluate the performance of SPChain. Thereafter we discuss whether SPChain can fulfill the requirements and prevent attacks proposed in Section 4. 
\subsection{Implementation}
In order to show the feasibility of SPChain, we build a Bitcoin network in a desktop computer (equipped with a Ubuntu 16.04 LTS, Intel (R) Core (TM) i5-8500 CPU of 3.00GHz and 16GB RAM). In our deployment of SPChain, we adopt a state-of-the-art Byzantine-resilient consensus protocol called BFT-SMaRt \cite{bessani2014state} which is widely used in different programs such as Hyperledger Fabric. We adopt the computing power of Bitcoin miners \footnote{\scriptsize{The computing power of the top 26 mining pools are 17.02\%, 15.9\%, 12.99\%, 11.25\%, 6.37\%, 6.15\%, 5.22\%, 4.58\%, 4.01\%, 3.81\%, 3.21\%, 1.84\%, 1.68\%, 1.64\%, 1.55\%, 1.20\%, 0.50\%, 0.44\%, 0.33\%, 0.1\%, 0.05\%, 0.04\%, 0.04\%, 0.03\%, 0.01\%, 0.01\% respectively. https://btc.com/stats/pool?pool mode=year}} in the real world to simulate the mining process of different medical institutions. We make use of Github programs Repuify \footnote{\scriptsize{https://github.com/JeevanPillay/repuify}} to simulate the communication process of SPChain. Similar to Bitcoin, we stipulate that keyblocks are generated every 10 minutes on average. Differently, the medical institutions which create microblocks send them at the highest rate they can maintain. 
\subsection{Performance evaluation}
We evaluate the performance of SPChain in the following aspects.
\begin{figure}[h]
	\includegraphics[width=10cm,height=7cm]{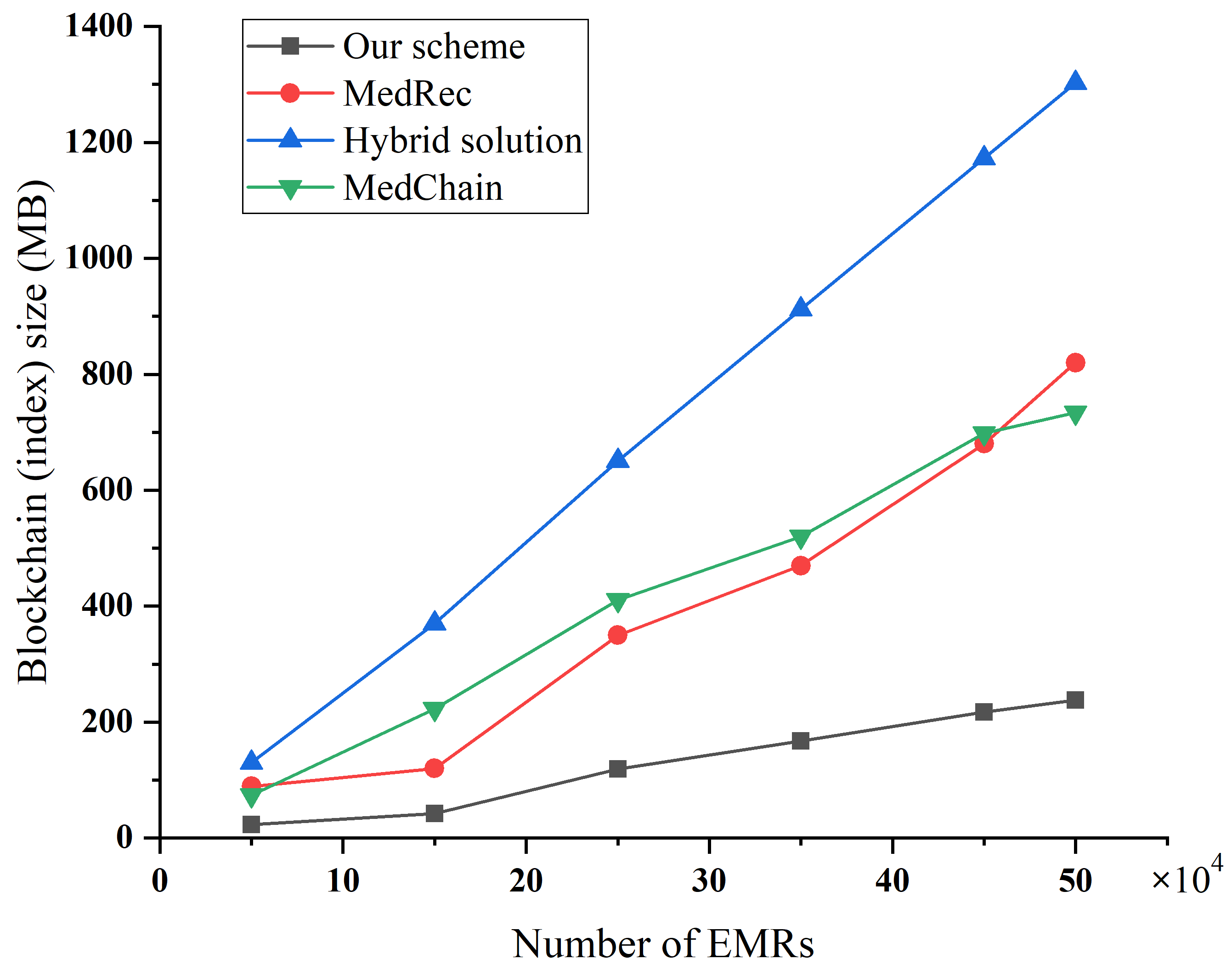}
	\caption{Storage overhead Comparison.}
	\label{fig11}
\end{figure}
\begin{table}
	\centering
	\caption{The size of different transactions.}
	\label{table3}
	\begin{tabular}{c|ccccc|c}
		\toprule
		\diagbox  {Type}{Size(byte)} & $\mathbf{type}$ & $\mathbf{input}$ & $\mathbf{output}$  & $\mathbf{data}$ & $\mathbf{sig}$&$\mathbf{total}$ \\
		\hline
		$T_R$ & 4 & 110 & 70 & 32 & 70 &286  \\
		$T_M$ & 4 & 110 & 70 & 288 & 70 &542  \\
		$T_L$ & 4 & 110 & 70 &  320&  70& 574\\
		\bottomrule
	\end{tabular}
\end{table}

\textbf{Storage cost.} We choose the Nursery dataset from University of California, Irvine (UCI) Machine Learning Repository \cite{dataset} as the test dataset. The dataset contains 12960 EMRs and the average size of each EMR is 32KB. We expand the number of EMRs in the original dataset to 500000 to evaluate SPChain. We leverage the chameleon hash function in Github \footnote{https://github.com/julwil/chameleon\_hash} to generate chameleon hash values for encrypted EMRs. Table \ref{table3} describes the sizes of different transactions. We also test the storage overhead of different schemes under different numbers of EMRs, and the results are shown in Fig. \ref{fig11}. Compared with hybrid solution \cite{yang2015hybrid}, MedRec \cite{azaria2016medrec} and MedChain \cite{shen2019medchain}, SPChain has lower storage overhead. For example, when sharing 250000 EMRs, the blockchain size of SPChain is 119MB, less than that of hybrid solution (350MB), MedChain (651MB) and MedRec (411MB).
\begin{figure}[htbp]
	\centering
	\begin{minipage}[t]{0.48\textwidth}
		\centering
		\includegraphics[width=6cm]{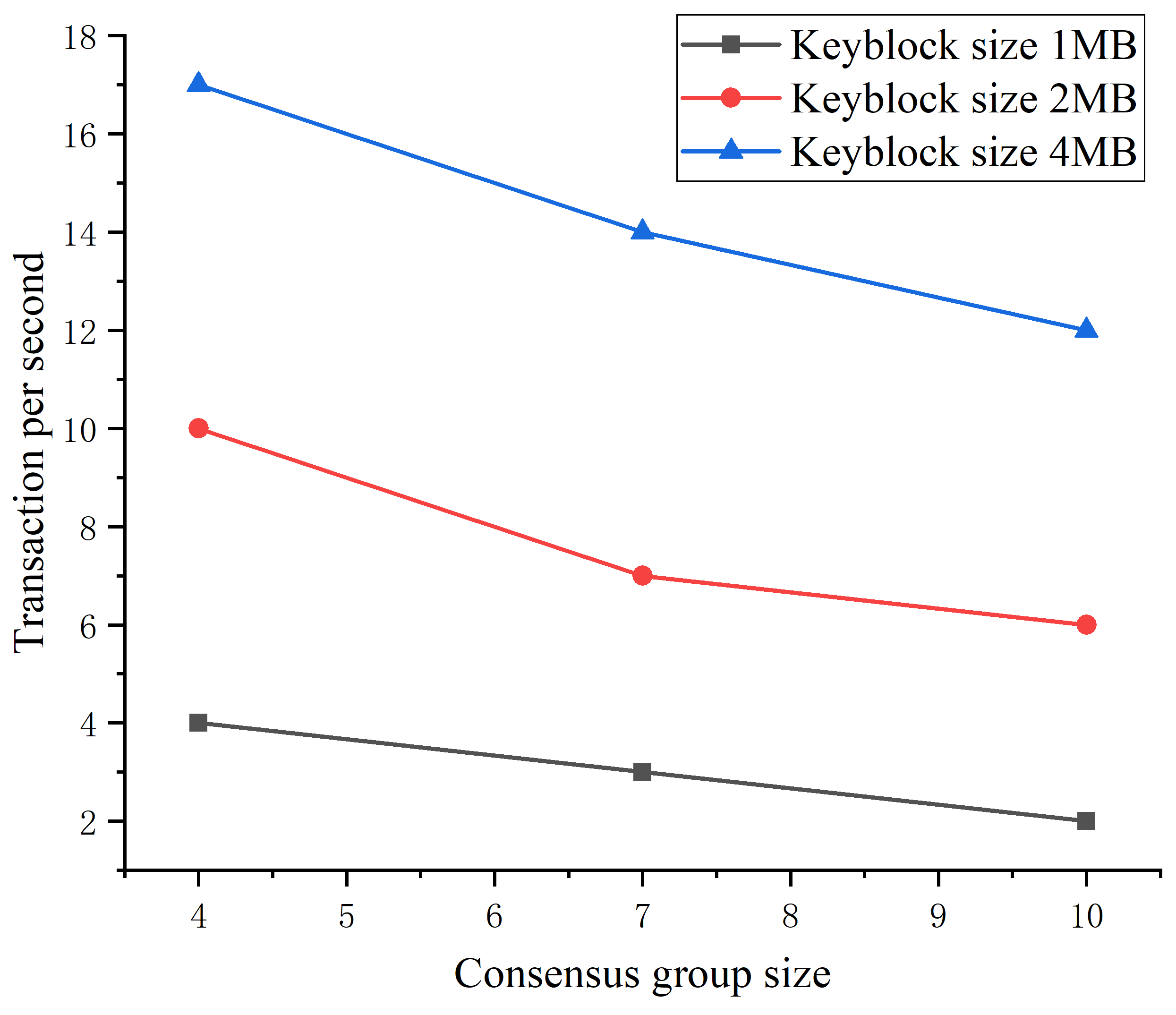}
		\caption{The throughput of SPChain (keyblocks).}
		\label{fig12}
	\end{minipage}
	\begin{minipage}[t]{0.48\textwidth}
		\centering
		\includegraphics[width=6cm]{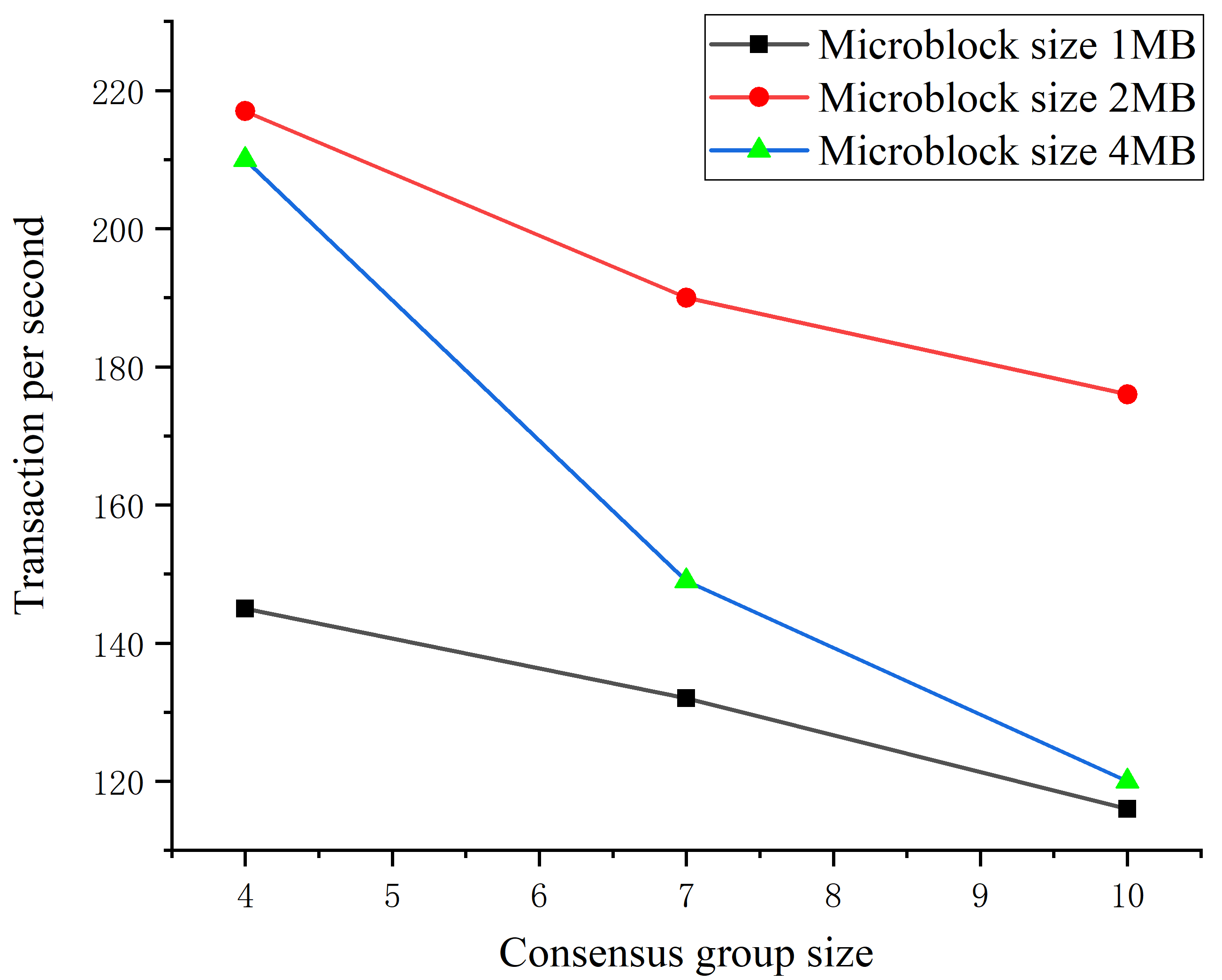}
		\caption{ The throughput of SPChain (microblocks).}
		\label{fig13}
	\end{minipage}
\end{figure}

\textbf{Throughput.} In this part we analyze the maximum throughput of our system. We assume the consensus group controls 90\% computing power. Since keyblocks in SPChain contain register transactions, so we analyze the throughput in terms of keyblocks and microblocks. From Fig. \ref{fig12} we can see that the throughput of keyblocks is similar to that of bitcoin systems since they are under the same mining strategy. When fix the block size to 4MB, the system has higher throughput than that of 1MB and 2MB. As for microblocks, our results in Fig. \ref{fig13} shows that when the block size is fixed, as the number of consensus nodes increases, throughput decreases gradually. For example, with the block size 1MB, the through decreases from 145 TPS to 116 TPS. Besides, from Fig. \ref{fig13} we can see that when fix the block size to 2MB, the system has higher throughput than that of 1MB and 4MB. In particular, when the consensus group consists of 4 nodes, the through can reach 218 TPS.
\begin{figure}[h]
	\includegraphics[height=8cm,width=11cm]{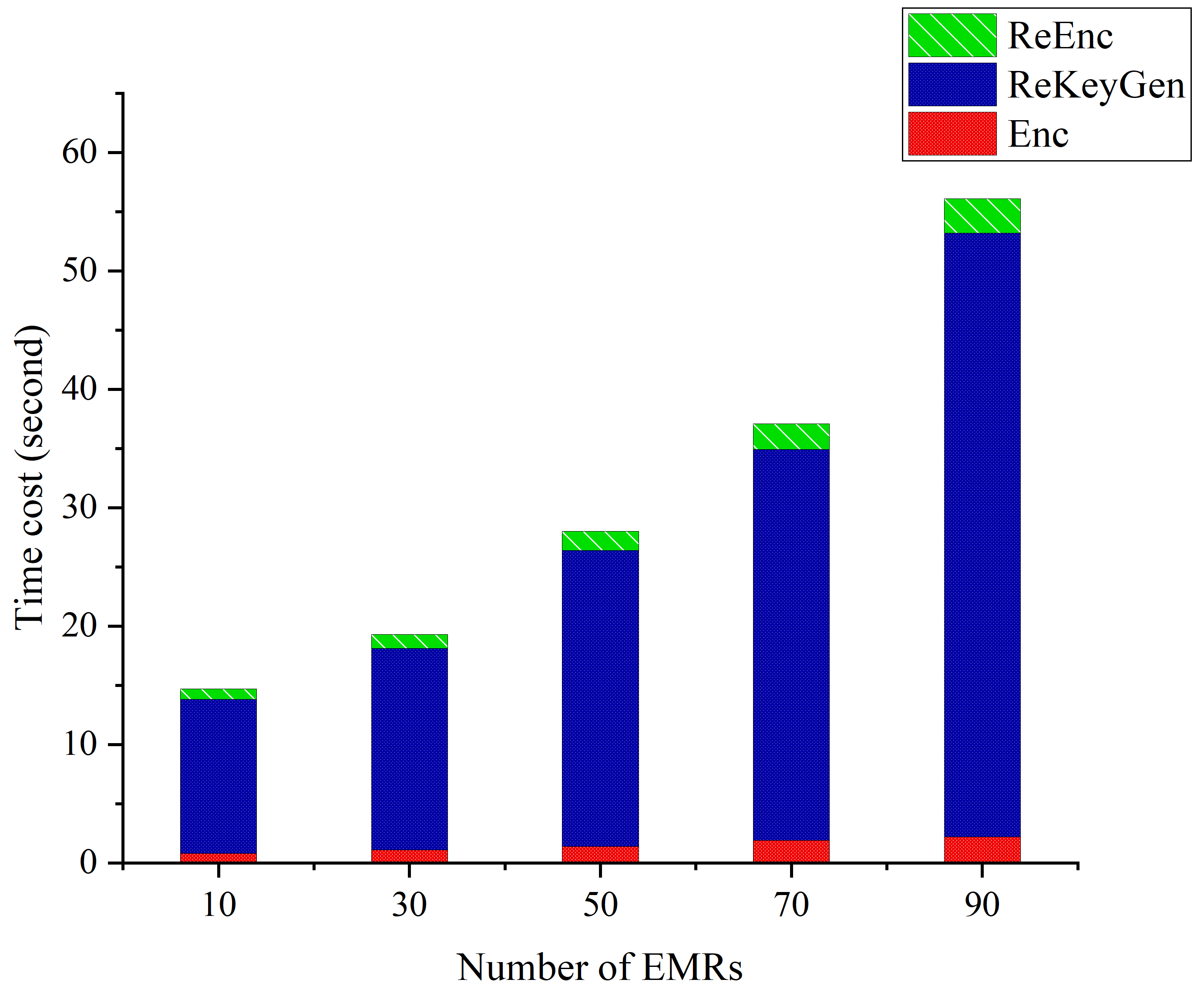}
	\caption{The time cost of EMRs sharing in a medical institution.}
	\label{fig14}
\end{figure}

\textbf{Time cost.} We analyze time cost in terms of EMRs sharing and EMRs retrieval. Note that patients encrypt EMRs before storing them to the local database of a medical institution and later the medical institution re-encrypts EMRs for sharing them. Fig. \ref{fig14} illustrates the data sharing time cost in SPChain. For example, when a patient shares 50 EMRs from medical institution $\tilde{A}$ to medical institution $\tilde{D}$, $\tilde{A}$ spends 21 seconds to generate the re-encryption key and 5 seconds to re-encrypt the shared EMRs.
\begin {table*}[h]
\centering\caption{Comparison between existing blockchain-based eHealth systems and SPChain.}
\label{tab4}
\resizebox{\textwidth}{!}{
	\small
	\begin{tabular}{c|c|c|c|c|c}
		\hline
		Scheme &\cite{xia2017bbds}&\cite{chen2019blockchain}&\cite{azaria2016medrec}&\cite{huang2020blockchain} & SPChain \\
		\hline
		First part (block)&$\mathsf{O(n)}$&$\mathsf{O(n)}$ &$\mathsf{O(n)}$& $\mathsf{O(n)}$ &  $\mathsf{O(1)}$ \\
		\hline
		Second part (record)&$\mathsf{O(logn)}$&$\mathsf{O(logn)}$ &$\mathsf{O(logn)}$&$\mathsf{O(logn)}$&$\mathsf{O(logn)}$  \\
		\hline
\end{tabular}}
\end{table*}

Table \ref{tab4} illustrates the time complexity of retrieving medical records in different schemes. In SPChain, the time consumption of retrieving an EMR is divided into two part. The first part is to locate a specific block. Due to the special construction of blocks in SPChain, each patient holds his/her own unique block. Thus a patient can acquire the block according to the block number directly, so the time complexity is $\mathsf{O(1)}$. The second part of time consumption is to locate specific medical records or subtrees by searching in Merkle tree which stores numerous EMRs. The average time complexity is $\mathsf{O(logn)}$. From Table \ref{tab4} we can conclude that our scheme performs better than other schemes in terms of retrieving a specific block.

\subsection{Security analysis}
SPChain can satisfy all the security requirements described in Section 4.3, according to the proposed transactions, blocks and chain structure.

\textbf{Confidentiality.} Note that the EMRs are encrypted by patients' PRE public-keys and stored in the local databases of medical institutions. In this circumstance, the medical institutions cannot acquire any information of the EMRs. When sharing the EMRs to other medical institutions, the encrypted EMRs can only be decrypted by the authorized medical institutions, which ensures that the patients' EMRs are not disclosed to unauthorized medical institutions during the sharing process. Thus SPChain guarantees the confidentiality of patients' EMRs.

\textbf{Patient centric sharing.} We use proxy re-encryption schemes to encrypt the patients' EMRs. In this case, medical institutions cannot use or share EMRs without patients' authorization because only the authorized medical institutions can decrypt the encrypted EMRs. Besides, only the reputable medical institutions have the right to request EMRs from patients.

\textbf{Quick retrieval.} We devise special block structure and chain structure for patients. As illustrated in table \ref{tab4}, SPChain has less time complexity in retrieval than other schemes.
\begin{figure}[h]
	\centering
	\includegraphics[height=7.5cm,width=11cm]{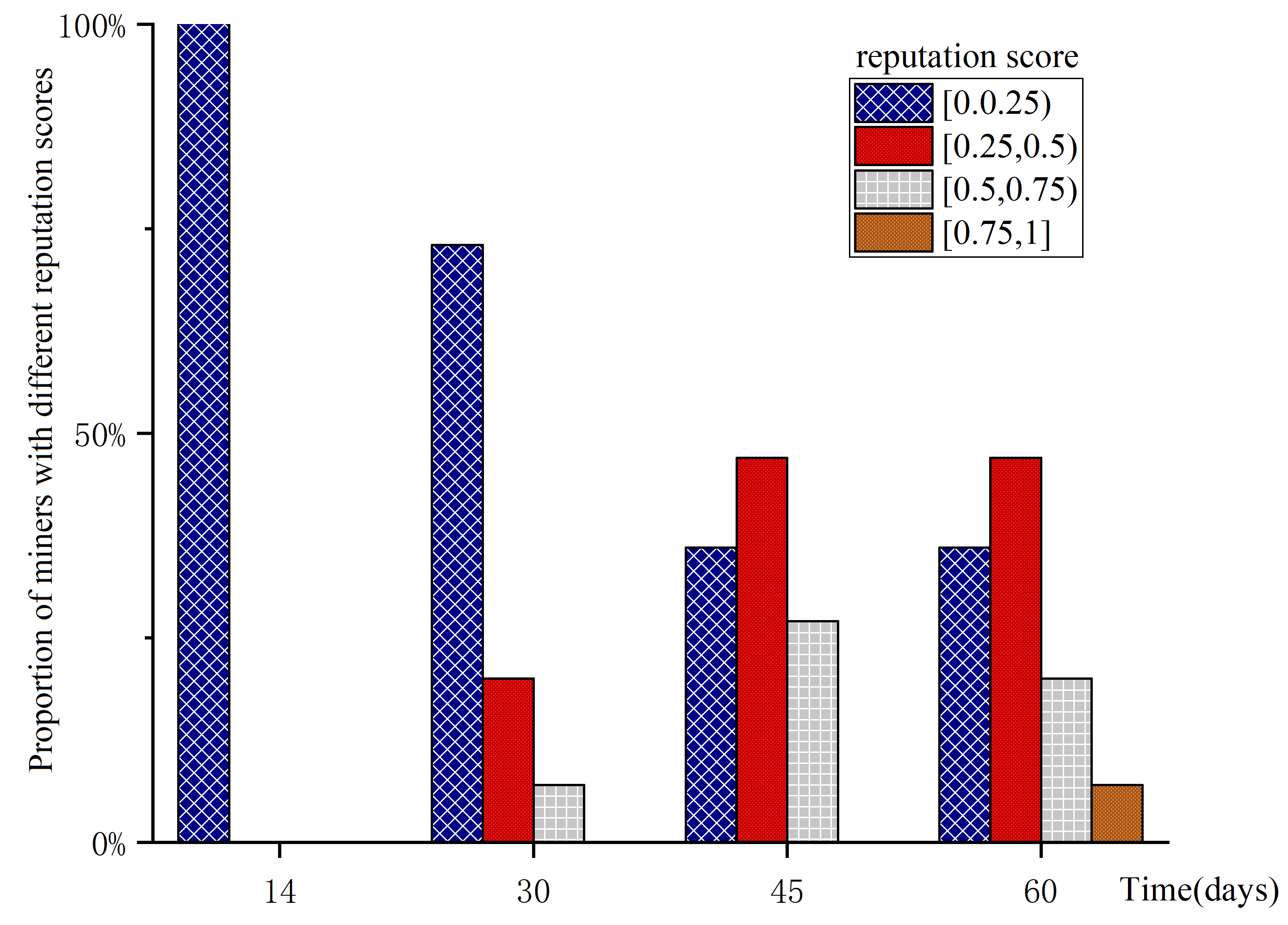}
	\caption{The distribution of medical institutions' reputation scores over time.}
	\label{fig15}
\end{figure}

\textbf{Label and correctness.} SPChain permits medical institutions to label wrong EMRs and allows patients to verify the correctness of the labeled EMRs. In the case of misdiagnosis, patients can ask authorized medical institutions to label the wrong EMRs with the label transaction and to generate the correct EMRs corresponding to the label transaction. With the parameter $r$, patients can invoke $\mathcal{CH}.\mathtt{HVer}$ to check whether the label is correct.

\textbf{Resistance to blockchain attacks.} To demonstrate that SPChain is resistant to blockchain attacks and SPChain attacks illustrated in Section 4.3, we simulate the reputation cores of medical institutions by choosing the top 26 mining pools in Bitcoin. We set the parameter $a=5000$ and $\lambda=20000$. Fig. \ref{fig15} describes the distribution of medical institutions with different reputation scores over time. With the operation of SPChain, the reputation score of miners increase gradually, and the higher computing power the miner holds, the higher the reputation score it receives.
\begin{figure}[h]
	\centering
	\includegraphics[height=8cm,width=10cm]{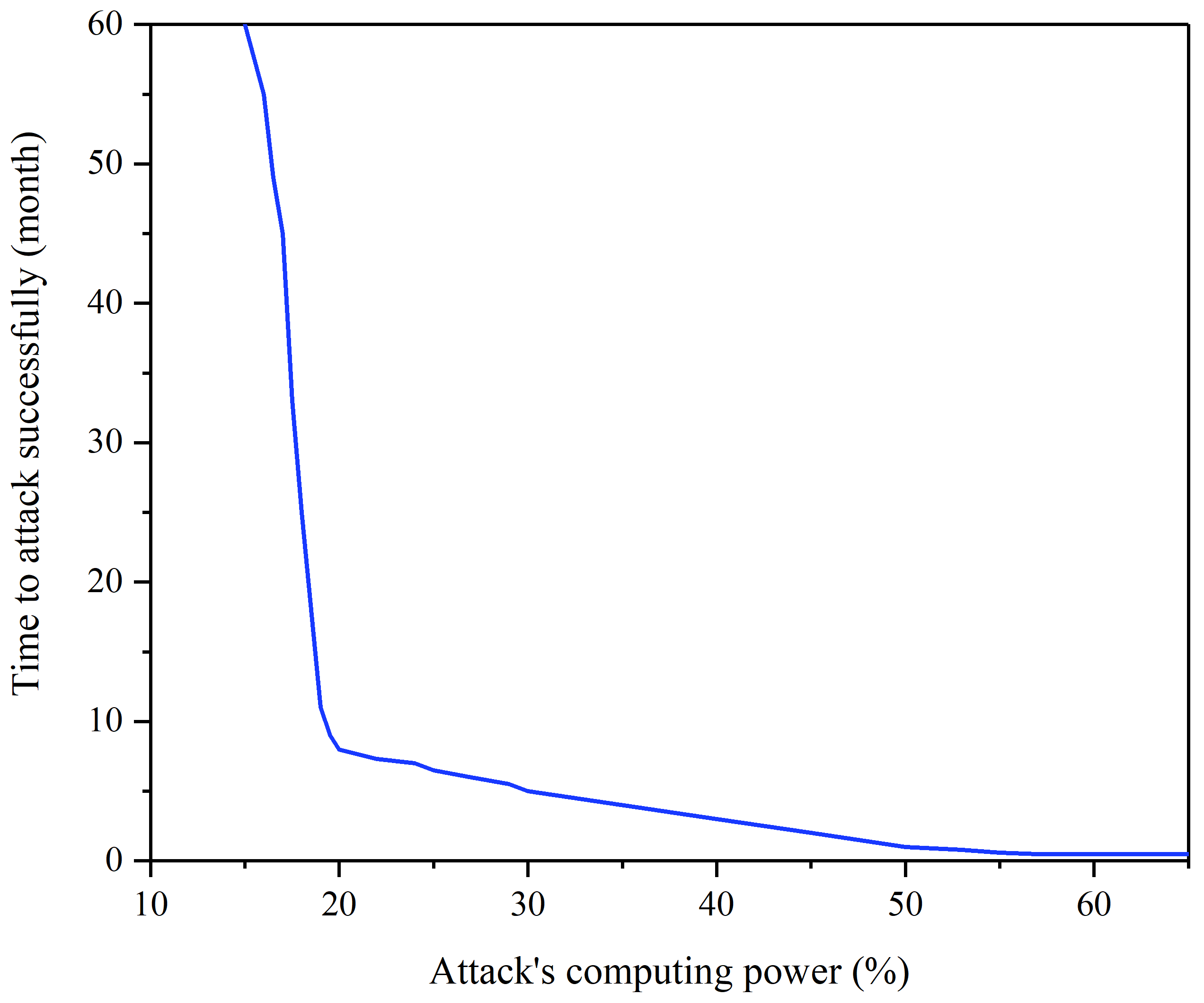}
	\caption{Time to break SPChain.}
	\label{fig16}
\end{figure}

\begin{itemize}
	\item \textit{51\% attacks and flash attacks.} SPChain is resilient to 51\% attacks and flash attacks. Fig \ref{fig16} shows the time required for an attacker with different computing powers to destroy SPChain. For example, an attacker with 65\% computing power would take a month to break SPChain. But after launching the attacks successfully, the attacker lose all his/her reputation scores he/she accumulates within one month. We point out that the longer the system runs, the longer it takes for the attacker to attack. After a sufficiently long period of time (a.k.a one year), SPChain can tolerate 51\% attacks and flash attacks.
	\item \textit{Selfish mining attacks.} SPChain pins each keyblock, and the pinned keyblocks cannot be rolled back. Every new created keyblock is chained behind the pinned keyblocks. So if an attacker publishes a keyblock which is conflict with the pinned keyblock, he/she cannot get advantage of gaining rewards over the honest miners because the keyblock he/she publishes would not be admitted by the system. When mining a new keyblock, the miners need to take the hash value of the previous keyblock and the last microblock of the penultimate keyblock as inputs. Since there is no conflict when generating microblocks, we do not need to consider the microblock withholding attacks. In summary, SPChain can resist selfish mining attacks.
	\item \textit{Sybil attacks.} Since SPChain adopts PoW consensus mechanism, all medical institutions in SPChain need a lot of computations to accumulate reputation scores to get voting rights. Thus SPChain can resist Sybil attacks.
\end{itemize}

\textbf{Resistance to SPChain attacks.} In this part we illustrate how SPChain can thwarts the following attacks.
\begin{itemize}
\item \textit{Reputation fraud attacks.} To prevent malicious medical institutions from creating fake patients to increase their reputation scores, we require patients to register in SPChain with the hash values of unique identifiers (e.g. ID numbers), thus the zombie patients can be filtered out of SPChain. Besides, we stipulate that every transaction should contain transaction fees, which can increase the attacker's cost to launch reputation fraud attacks.
\item  \textit{Inhibition attacks.} As shown in Algorithm \ref{alg:2}, the medical institution in the consensus group should deal with transactions according to the medical institutions' reputation scores, which can prevent an attacker from launching inhibition attacks.
\end{itemize}

Finally we compare SPChain with other blockchain-based eHealth systems. From Table \ref{tab5} we can see that \cite{xia2017bbds}, \cite{chen2019blockchain}, \cite{azaria2016medrec} and \cite{shen2019medchain} cannot label wrong EMRs in the case of misdiagnosis. And these systems are vulnerable to blockchain underlying attacks such as flash attacks and selfish mining attacks. 
\begin {table*}[h]
\centering\caption{Comparison between existing systems and SPChain.}
\label{tab5}
\resizebox{\textwidth}{!}{
	\small
	\begin{tabular}{c|cccccc}
		\hline
		\bf{Scheme} &\bf{Share}&\bf{Privacy} &\bf{Label wrong EMRs}& \bf{Flash attacks}& \bf{Selfish mining attacks}&\bf{Sybil attacks} \\
		\hline
		\cite{xia2017bbds} &$\checkmark$&$\checkmark$ &$\times$& $\times$& $\times$&  $\times$\\
		\hline
		\cite{chen2019blockchain} &$\checkmark$&$\checkmark$ &$\times$& $\times$& $\times$& $\checkmark$ \\
		\hline
		\cite{azaria2016medrec} &$\checkmark$&$\times$ &$\times$& $\times$& $\times$ & $\checkmark$\\
		\hline
		\cite{shen2019medchain} &$\checkmark$&$\checkmark$ &$\times$& $\times$& $\times$& $\checkmark$\\
		\hline
		\bf{SPChain} &$\checkmark$&$\checkmark$ &$\checkmark$& $\checkmark$& $\checkmark$& $\checkmark$ \\
		\hline
\end{tabular}}
\end{table*}
\section{Conclusion and future work}
In this paper, we achieved medical data sharing and retrieval in a privacy-preserving manner. We have designed register transactions, medical transactions and label transactions for patients to achieve register, EMR uploading and labeling wrong EMRs respectively. To quickly retrieve the patients' medical history, we devised special microblocks for each patient. We also designed an reputation-based consensus mechanism to motivate medical institutions such that the medical institutions can participate in the mining process to accumulate reputation scores to require patients' EMRs. Finally, we discussed how the proposal can satisfy the security requirements by developing the system in an analog network with the miner distribution data in the real world. We compared the time costs of data retrieval and the storage costs with some existing solutions. The results proved the feasibility and effectiveness of our system. We also analyzed the impact of our reputation-based consensus mechanism on the security of SPChain and demonstrated that SPChain can effectively resist blockchain attacks and the proposed SPChain attacks.

For the future work, we intend to reduce the patients' communication overhead and improve the throughput of our system.
\section*{References}

\bibliography{SPChain}

\end{document}